\documentclass[12pt]{article}
\usepackage{graphicx}
\def\ffrac#1#2{\textstyle{#1\over#2}\displaystyle}

\begin{document}
\pagestyle{myheadings}
\parindent 0mm
\parskip 6pt
\markright{CFT and Statistical Mechanics}

\title{Conformal Field Theory and Statistical Mechanics\footnote{Lectures given at the
Summer School on {\sl Exact methods in low-dimensional statistical
physics and quantum computing}, les Houches, July 2008.}}

\author{John Cardy}
%
\date{July 2008}
\maketitle
%

%
\newpage
\tableofcontents
\section{Introduction}
It is twenty years ago almost to the day that the les Houches
school on {\sl Fields, Strings and Critical Phenomena} took place.
It came on the heels of a frenzied period of five years or so
following the seminal paper of Belavin, Polyakov and Zamolodchikov
(BPZ) in which the foundations of conformal field theory and
related topics had been laid down, and featured lectures on CFT by
Paul Ginsparg, Ian Affleck, Jean-Bernard Zuber and myself, related
courses and talks by Hubert Saleur, Robert Dijkgraaf, Bertrand
Duplantier, Sasha Polyakov, Daniel Friedan, as well as lectures on
several other topics. The list of young participants, many of whom
have since gone on to make their own important contributions, is
equally impressive.

Twenty years later, CFT is an essential item in the toolbox of
many theoretical condensed matter physicists and string theorists.
It has also had a marked impact in mathematics, in algebra,
geometry and, more recently, probability theory. It is the purpose
of these lectures to introduce some of these tools. In some ways
they are an updated version of those I gave in 1988. However,
there are some important topics there which, in order to include
new material, I will omit here, and I would encourage the diligent
student to read both versions. I should stress that the lectures
will largely be about Conformal Field Theor\underline{y}, rather
than Conformal Field Theor\underline{ies}, in the sense that I'll
be describing rather generic properties of CFT rather than
discussing particular examples. That said, I don't want the
discussion to be too abstract, and in fact I will have a very
specific idea about what are the kind of CFTs I will be
discussing: the scaling limit (in a sense to be described) of
critical lattice models (either classical, in two dimensions, or
quantum in 1+1 dimensions.) This will allow us to have, I hope,
more concrete notions of the mathematical objects which are being
discussed in CFT. However there will be no attempt at mathematical
rigour. Despite the fact that CFT can be developed axiomatically,
I think that for this audience it is more important to understand
the physical origin of the basic ideas.

\section{Scale invariance and conformal invariance in critical behaviour}
\subsection{Scale invariance}
The prototype lattice model we have at the back of our minds is
the ferromagnetic Ising model. We take some finite domain $\cal D$
of $d$-dimensional euclidean space and impose a regular lattice,
say hypercubic, with lattice constant $a$. With each node $r$ of
the lattice we associate a binary-valued spin $s(r)=\pm1$. Each
configuration $\{s\}$ carries a relative weight
$W(\{s\})\propto\exp(\sum_{rr'\in{\cal D}}J(r-r')s(r)s(r))$ where
$J(r-r')>0$ is some short-ranged interaction (i.e. it vanishes if
$|r-r'|$ is larger than some fixed multiple of $a$.)

The usual kinds of local observable $\phi^{\rm lat}_j(r)$
considered in this lattice model are sums of products of nearby
spins over some region of size $O(a)$, for example the local
magnetisation $s(r)$ itself, or the energy density
$\sum_{r'}J(r-r')s(r)s(r')$. However it will become clear later on
that there are other observables, also labelled by a single point
$r$, which are functions of the whole configuration $\{s\}$. For
example, we shall show that this Ising model can be mapped onto a
gas of non-intersecting loops. An observable then might depend on
whether a particular loop passes through the given point $r$. From
the point of view of CFT, these observables are equally valid
objects.

Correlation functions, that is expectation values of products of
local lattice observables, are then given by
$$
\langle\phi^{\rm lat}_1(r_1)\phi^{\rm lat}_2(r_2)\ldots\phi^{\rm
lat}_n(r_n)\rangle=Z^{-1}\sum_{\{s\}}\phi^{\rm
lat}_1(r_1)\ldots\phi^{\rm lat}_n(r_n)W(\{s\})\,,
$$
where $Z=\sum_{\{s\}}W(\{s\})$ is the partition function. In
general, the connected pieces of these correlations functions fall
off over the same distance scale as the interaction $J(r-r')$,
but, close to a critical point, this correlation length $\xi$ can
become large, $\gg a$.

The \em scaling limit \em is obtained by taking $a\to0$ while
keeping $\xi$ and the domain $\cal D$ fixed. In general the
correlation functions as defined above do not possess a finite
scaling limit. However, the theory of renormalisation (based on
studies in exactly solved models, as well as perturbative analysis
of cut-off quantum field theory) suggests that in general there
are particular linear combinations of local lattice observables
which are \em multiplicatively renormalisable\em. That is, the
limit
\begin{equation}
\label{eq1}
 \lim_{a\to0}a^{-\sum_{j=1}^nx_j}
\langle\phi^{\rm lat}_1(r_1)\phi^{\rm lat}_2(r_2)\ldots\phi^{\rm
lat}_n(r_n)\rangle
\end{equation}
exists for certain values of the $\{x_j\}$. We usually denote this
by
\begin{equation}
\label{eq2}
\langle\phi_1(r_1)\phi_2(r_2)\ldots\phi_n(r_n)\rangle\,,
\end{equation}
and we often think of it as the expectation value of the product
the random variables $\phi_j(r_j)$, known as \em scaling fields
\em (sometimes scaling operators, to be even more confusing) with
respect to some `path integral' measure. However it should be
stressed that this is only an occasionally useful fiction, which
ignores all the wonderful subtleties of renormalised field theory.
The basic objects of QFT are the correlation functions. The
numbers $\{x_j\}$ in (\ref{eq1}) are called the \em scaling
dimensions\em.

One important reason why this is not true in general is that the
limit in (\ref{eq1}) in fact only exists if the points $\{r_j\}$
are non-coincident. The correlation functions in (\ref{eq2}) are
singular in the limits when $r_i\to r_j$. However, the nature of
these singularities is prescribed by the \em operator product
expansion \em (OPE)
\begin{equation}
\label{ope1}
\langle\phi_i(r_i)\phi_j(r_j)\ldots\rangle=\sum_kC_{ijk}(r_i-r_j)
\langle\phi_k((r_i+r_j)/2)\ldots\rangle\,.
\end{equation}
The main point is that, in the limit when $|r_i-r_j|$ is much less
than the separation between $r_i$ and all the other arguments in
$\ldots$, the coefficients $C_{ijk}$ are independent of what is in
the dots. For this reason, (\ref{ope1}) is often written as
\begin{equation}
\label{ope2} \phi_i(r_i)\cdot\phi_j(r_j)=\sum_kC_{ijk}(r_i-r_j)
\phi_k((r_i+r_j)/2)\,,
\end{equation}
although it should be stressed that this is merely a short-hand
for (\ref{ope1}).

So far we have been talking about how to get a continuum
(euclidean) field theory as the scaling limit of a lattice model.
In general this will be a massive QFT, with a mass scale given by
the inverse correlation length $\xi^{-1}$. In general, the
correlation functions will depend on this scale. However, at a
(second-order) critical point the correlation length $\xi$
diverges, that is the mass vanishes, and there is no length scale
in the problem besides the overall size $L$ of the domain $\cal
D$.

The fact that the scaling limit of (\ref{eq1}) exists then implies
that, instead of starting with a lattice model with lattice
constant $a$, we could equally well have started with one with
some fraction $a/b$. This would, however, be identical with a
lattice model with the original spacing $a$, in which all lengths
(including the size of the domain $\cal D$) are multiplied by $b$.
This implies that the correlation functions in (\ref{eq2}) are \em
scale covariant\em:
\begin{equation} \label{eq3}
\langle\phi_1(br_1)\phi_2(br_2)\ldots\phi_n(br_n)\rangle_{b{\cal
D}}=b^{-\sum_jx_j}\langle\phi_1(r_1)\phi_2(r_2)\ldots\phi_n(r_n)\rangle_{\cal
D}\,.
\end{equation}
Once again, we can write this in the suggestive form
\begin{equation}\label{eq5}
\phi_j(br)=b^{-x_j}\phi_j(r)\,,
\end{equation}
as long as what we really mean is (\ref{eq3}).

In a massless QFT, the form of the OPE coefficients in
(\ref{ope2}) simplifies: by scale covariance
\begin{equation}\label{ope3}
C_{ijk}(r_j-r_k)=\frac{c_{ijk}}{|r_i-r_j|^{x_i+x_j-x_k}}\,,
\end{equation}
where the $c_{ijk}$ are pure numbers, and \em universal \em if the
2-point functions are normalised so that
$\langle\phi_j(r_1)\phi_j(r_2)\rangle=|r_1-r_2|^{-2x_j}$. (This
assumes that the scaling fields are all rotational scalars --
otherwise it is somewhat more complicated, at least for general
dimension.)

From scale covariance, it is a simple but powerful leap to \em
conformal covariance\em: suppose that the scaling factor $b$ in
(\ref{eq3}) is a slowly varying function of position $r$. Then we
can try to write a generalisation of (\ref{eq3}) as
\begin{equation}
\label{conf}
\langle\phi_1(r_1')\phi_2(r_2')\ldots\phi_n(r_n')\rangle_{{\cal
D}'}=\prod_{j=1}^nb(r_j)^{-x_j}\langle\phi_1(r_1)\phi_2(r_2)\ldots\phi_n(r_n)\rangle_{\cal
D}\,,
\end{equation}
where $b(r)=|\partial r'/\partial r|$ is the local jacobian of the
transformation $r\to r'$.

For what transformations $r\to r'$ do we expect (\ref{conf}) to
hold? The heuristic argument runs as follows: if the theory is
local (that is the interactions in the lattice model are
short-ranged), then as long as the transformations looks \em
locally \em like a scale transformation (plus a possible
rotation), then (\ref{conf}) may be expected to hold. (In
Sec.~\ref{secT} we will make this more precise, based on the
assumed properties of the stress tensor, and argue that in fact it
holds only for a special class of scaling fields $\{\phi_j\}$
called primary.)

It is most important that the underlying lattice does \em not \em
transform (otherwise the statement is a tautology): (\ref{conf})
relates correlation functions in $\cal D$, defined in terms of the
limit $a\to0$ of a model on a regular lattice superimposed on
$\cal D$, to correlation functions defined by a regular lattice
superimposed on ${\cal D}'$.

Transformations which are locally equivalent to a scale
transformation and rotation, that is, have no local components of
shear, also locally preserve angles and are called \em
conformal\em.

\subsection{Conformal mappings in general}
Consider a general infinitesimal transformation (in flat space)
$r^\mu\to {r'}^\mu=r^\mu+\alpha^\mu(r)$ (we distinguish upper and
lower indices in anticipation of using coordinates in which the
metric is not diagonal.) The shear component is the traceless
symmetric part
$$
\alpha^{\mu,\nu}+\alpha^{\nu,\mu}-(2/d){\alpha^{\lambda}}_{,\lambda}g^{\mu\nu}\,,
$$
all $\frac12d(d+1)-1$ components of which must vanish for the
mapping to be conformal. For general $d$ this is very restrictive,
and in fact, apart from uniform translations, rotations and scale
transformations, there is only one other type of solution
$$
\alpha^\mu(r)=b^\mu r^2-2(b\cdot r)r^\mu\,,
$$
where $b^\mu$ is a constant vector. These are in fact the
composition of the finite conformal mapping of inversion $r^\mu\to
r^{\mu}/|r|^2$, followed by an infinitesimal translation $b^\mu$,
followed by a further inversion. They are called the special
conformal transformations, and together with the others, they
generate a group isomorphic to SO$(d+1,1)$.

These special conformal transformations have enough freedom to fix
the form of the 3-point functions in ${\bf R}^d$ (just as scale
invariance and rotational invariance fixes the 2-point functions):
for scalar operators\footnote{The easiest way to show this it to
make an inversion with an origin very close to one of the points,
say $r_1$, and then use the OPE, since its image is then very far
from those of the other two points.}
\begin{equation}\label{3pt}
\langle\phi_1(r_1)\phi_2(r_2)\phi_3(r_3)\rangle=\frac{c_{123}}
{|r_1-r_2|^{x_1+x_2-x_3}|r_2-r_3|^{x_2+x_3-x_1}|r_3-r_1|^{x_3+x_1-x_2}}\,.
\end{equation}
Comparing with the OPE (\ref{ope2},\ref{ope3}), and assuming
non-degeneracy of the scaling dimensions\footnote{This and other
properties fail in so-called logarithmic CFTs.}, we see that
$c_{123}$ is the same as the OPE coefficient defined earlier. This
shows that the OPE coefficients $c_{ijk}$ are symmetric in their
indices.

In two dimensions, the condition that $\alpha^\mu(r)$ be conformal
imposes only two differential conditions on two functions, and
there is a much wider class of solutions. These are more easily
seen using \em complex coordinates\em\footnote{For many CFT
computations we may treat $z$ and $\bar z$ as independent,
imposing only at the end that they should be complex conjugates.}
$z\equiv r^1+ir^2$, $\bar z\equiv r^1-ir^2$, so that the line
element is $ds^2=dzd\bar z$, and the metric is
$$
g_{\mu\nu}=\left(\matrix{0&\ffrac12\cr \ffrac12&0}\right)\qquad
g^{\mu\nu}=\left(\matrix{0&2\cr 2&0}\right)\,.
$$
In this basis, two of the conditions are satisfied identically and
the others become
$$
\alpha^{z,z}=\alpha^{\bar z,\bar z}=0\,,
$$
which means that $\partial\alpha^z/\partial\bar
z=\partial\alpha^{\bar z}/\partial z=0$, that is, $\alpha^z$ is a
holomorphic function $\alpha(z)$ of $z$, and $\alpha^{\bar z}$ is
an antiholomorphic function.

Generalising this to a finite transformation, it means that
conformal mappings $r\to r'$ correspond to  functions $z\to
z'=f(z)$ which are analytic in $\cal D$. (Note that the only such
functions on the whole Riemann sphere are the M\"obius
transformations $f(z)=(az+b)/(cz+d)$, which are the finite special
conformal mappings.)

In passing, let us note that complex coordinates give us a nice
way of discussing non-scalar fields: if, for example, under a
rotation $z\to ze^{i\theta}$, $\phi_j(z,\bar z)\to
e^{is_j\theta}\phi_j$, we say that $\phi_j$ has conformal spin
$s_j$ (not related to quantum mechanical spin), and under a
combined transformation $z\to\lambda z$ where
$\lambda=be^{i\theta}$ we can write (in the same spirit as
(\ref{eq5}))
$$
\phi_j(\lambda z,\bar\lambda\bar z)=
\lambda^{-\Delta_j}{\bar\lambda}^{-\overline{\Delta}_j}\phi_j(z,\bar
z)\,,
$$
where $x_j=\Delta_j+\overline{\Delta}_j$,
$s_j=\Delta_j-\overline{\Delta}_j$.
$(\Delta_j,\overline{\Delta}_j)$ are called the complex scaling
dimensions of $\phi_j$ (although they are usually both real, and
not necessarily complex conjugates of each other.)

\section{The role of the stress tensor}\label{secT}
Since we wish to explore the consequences of conformal invariance
for correlation functions in a \em fixed \em domain $\cal D$
(usually the entire complex plane), it is necessary to consider
transformations which are \em not \em conformal everywhere. This
brings in the stress tensor $T_{\mu\nu}$ (also known as the
stress-energy tensor or the (improved) energy-momentum tensor). It
is the object appearing on the right hand side of Einstein's
equations in curved space. In a classical field theory, it is
defined in terns of the response of the action $S$ to a general
infinitesimal transformation $\alpha^\mu(r)$:
\begin{equation}\label{dS}
\delta S=-\frac1{2\pi}\int T_{\mu\nu}\alpha^{\mu,\nu}d^2r
\end{equation}
(the $(1/2\pi)$ avoids awkward such factors later on.) Invariance
of the action under translations and rotations implies that
$T_{\mu\nu}$ is conserved and symmetric. Moreover if $S$ is scale
invariant, $T_{\mu\nu}$ is also traceless. In complex coordinates,
the first two conditions imply that $T_{z\bar z}+T_{\bar zz}=0$
and $T_{z\bar z}=T_{\bar zz}$, so they both vanish, and the
conservation equations then read $\partial^zT_{zz}=2\partial
T_{zz}/\partial\bar z=0$ and $\partial T_{\bar z \bar z}/\partial
z=0$. Thus the non-zero components $T\equiv T_{zz}$ and $\overline
T\equiv T_{\bar z\bar z}$ are respectively holomorphic and
antiholomorphic fields. Now if we consider a more general
transformation for which $\alpha^{\mu,\nu}$ is symmetric and
traceless, that is a conformal transformation, we see that $\delta
S=0$ in this case also. Thus, at least classically, we see that
scale invariance and rotational invariance imply conformal
invariance of the action, at least if (\ref{dS}) holds. However if
the theory contains long-range interactions, for example, this is
no longer the case.

In a quantum 2d CFT, it is assumed that the above analyticity
properties continue to hold at the level of correlation functions:
those of $T(z)$ and $\overline T(\bar z)$ are holomorphic and
antiholomorphic functions of $z$ respectively (except at
coincident points.)

\subsubsection{An example - free (gaussian) scalar field}
The prototype CFT is the free, or gaussian, massless scalar field
$h(r)$ (we use this notation for reasons that will emerge later).
It will turn out that many other CFTs are basically variants of
this. The classical action is
$$
S[h]=(g/4\pi)\int(\partial_\mu h)(\partial^\mu h)d^2r\,.
$$
Since $h(r)$ can take any real value, we could rescale it to
eliminate the coefficient in front, but in later extensions this
will have a meaning, so we keep it. In complex coordinates,
$S\propto\int(\partial_zh)(\partial_{\bar z}h)d^2z$, and it is
easy to see that this is conformally invariant under $z\to
z'=f(z)$, since $\partial_z=f'(z)\partial_{z'}$, $\partial_{\bar
z}=\overline{f'(z)}\partial_{{\bar z}'}$ and
$d^2z=|f'(z)|^{-2}d^2z'$. This is confirmed by calculating
$T_{\mu\nu}$ explicitly: we find $T_{z\bar z}=T_{\bar zz}=0$, and
$$
T=T_{zz}=-g(\partial_zh)^2,\qquad\overline T=T_{\bar z\bar
z}=-g(\partial_{\bar z}h)^2\,.
$$
These are holomorphic
(resp.~antiholomorphic) by virtue of the classical equation of
motion $\partial_z\partial_{\bar z}h=0$.

In the quantum field theory, a given configuration $\{h\}$ is
weighted by $\exp(-S[h])$. The 2-point function is\footnote{this
is cut-off at small $k$ by the assumed finite size $L$, but we are
here also assuming that the points are far from the boundary.}
$$
\langle h(z,\bar z)h(0,0)\rangle=\frac{2\pi}g\int\frac{e^{ik\cdot
r}}{k^2}\frac{d^2k}{(2\pi)^2}\sim-(1/2g)\log(z\bar z/L^2)\,.
$$
This means that $\langle T\rangle$ is formally divergent. It can
be made finite, for example, by point-splitting and subtracting
off the divergent piece:
\begin{equation}\label{Tff}
T(z)=-g\lim_{\delta\to0}\left(\partial_zh(z+\ffrac12\delta)\partial_zh(z
-\ffrac12\delta)-\frac1{2g\delta^2}\right)\,.
\end{equation}
This doesn't affect the essential properties of $T$.

\subsection{Conformal Ward identity}\label{CWI}
\begin{figure}
\centering
\includegraphics[width=5cm]{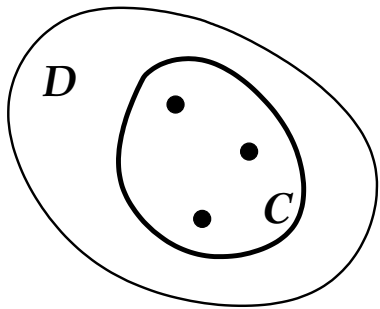}
\caption{\label{contour}\small We consider an infinitesimal
transformation which is conformal within $C$ and the identity in
the complement in $\cal D$. }
\end{figure}

Consider a general correlation function of scaling fields
$\langle\phi_1(z_1,\bar z_1)\phi_2(z_2,\bar
z_2)\ldots\rangle_{\cal D}$ in some domain $\cal D$. We want to
make an infinitesimal conformal transformation $z\to
z'=z+\alpha(z)$ on the points $\{z_j\}$, without modifying $\cal
D$. This can be done by considering a contour $C$ which encloses
all the points $\{z_j\}$ but which lies wholly within $\cal D$,
such that the transformation is conformal within $C$, and the
identity $z'=z$ outside $C$ (Fig.~\ref{contour}).
This gives rise to an (infinitesimal) discontinuity on $C$, and,
at least classically to a modification of the action $S$ according
to (\ref{dS}). Integrating by parts, we find $\delta
S=(1/2\pi)\int_C T_{\mu\nu}\alpha^\mu n^\nu d\ell$, where $n^\nu$
is the outward-pointing normal and $d\ell$ is a line element of
$C$. This is more easily expressed in complex coordinates, after
some algebra, as
$$
\delta S=\frac1{2\pi i}\int_C\alpha(z)T(z)dz+\mbox{complex
conjugate}\,.
$$
This extra factor can then be expanded, to first order in
$\alpha$, out of the weight $\exp(-S[h]-\delta S)\sim(1-\delta
S)\exp(-S[h])$, and the extra piece $\delta S$ considered as an
insertion into the correlation function. This is balanced by the
explicit change in the correlation function under the conformal
transformation:
\begin{equation}\label{dC}
\delta\langle\phi_1(z_1,\bar z_1)\phi_2(z_2,\bar
z_2)\ldots\rangle=\frac1{2\pi i}\int_C\alpha(z) \langle
T(z)\phi_1(z_1,\bar z_1)\phi_2(z_2,\bar z_2)\ldots\rangle dz+
\mbox{c.c.}
\end{equation}

Let us first consider the case when $\alpha(z)=\lambda(z-z_1)$,
that is corresponds to a combined rotation and scale
transformation. In that case
$\delta\phi_1=(\Delta_1\lambda+\overline{\Delta}_1\bar\lambda)\phi_1$,
and therefore, equating coefficients of $\lambda$ and
$\bar\lambda$,
$$
\int_C(z-z_1)\langle T(z)\phi_1(z_1,\bar z_1)\phi_2(z_2,\bar
z_2)\ldots\rangle \frac{dz}{2\pi i}=\Delta_1
\langle\phi_1(z_1,\bar z_1)\phi_2(z_2,\bar
z_2)\ldots\rangle+\cdots\,.
$$
Similarly, if we take $\alpha=$ constant, corresponding to a
translation, we have $\delta\phi_j\propto\partial_{z_j}\phi_j$, so
$$
\int_C \langle T(z)\phi_1(z_1,\bar z_1)\phi_2(z_2,\bar
z_2)\ldots\rangle \frac{dz}{2\pi i}=\sum_j\Delta_j\partial_{z_j}
\langle\phi_1(z_1,\bar z_1)\phi_2(z_2,\bar z_2)\ldots\rangle\,.
$$
Using Cauchy's theorem, these two equations tell us about two of
the singular terms in the OPE of $T(z)$ with a general scaling
field $\phi_j(z_j,\bar z_j)$:
\begin{equation}\label{Tope}
T(z)\cdot\phi_j(z_j,\bar
z_j)=\cdots+\frac{\Delta_j}{(z-z_j)^2}\phi_j(z_j,\bar z_j)
+\frac1{z-z_j}\partial_{z_j}\phi_j(z_j,\bar z_j)+\cdots\,.
\end{equation}
Note that only integer powers can occur, because the correlation
function is a meromorphic function of $z$.

\noindent{\em Example of the gaussian free field.\em}  If we take
$\phi^{\rm lat}_q(r)=e^{iqh(r)}$ then
$$
\langle\phi^{\rm lat}_q(r_1)\phi^{\rm lat}_{-q}(r_2)\rangle=
\exp\big(-\ffrac12q^2\langle(h(r_1)-h(r_2))^2\rangle\big)\sim
\left(\frac a{|r_1-r_2|}\right)^{q^2/g}\,,
$$
which means that the renormalised field $\phi_q\sim
a^{-q^2/2g}\phi_q^{\rm lat}$ has scaling dimension $x_q=q^2/2g$.
It is then a nice exercise in Wick's theorem to check that the OPE
with the stress tensor (\ref{Tope}) holds with $\Delta_q=x_q/2$.
(Note that in this case the multiplicative renormalisation of
$\phi_q$ is equivalent to ignoring all Wick contractions between
fields $h(r)$ at the same point.)

Now suppose each $\phi_j$ is such that the terms
$O\big((z-z_j)^{-2-n}\big)$ with $n\geq1$ in (\ref{Tope}) are
absent. Since a meromorphic function is determined entirely by its
singularities, we then know the correlation function $\langle
T(z)\ldots\rangle$ exactly:
\begin{equation}\label{WI}
\langle T(z)\phi_1(z_1,\bar z_1)\phi_2(z_2,\bar z_2)\ldots\rangle=
\sum_j\left(\frac{\Delta_j}{z-z_j)^2}
+\frac1{z-z_j}\partial_{z_j}\right)\langle\phi_1(z_1,\bar
z_1)\phi_2(z_2,\bar z_2)\ldots\rangle\,.
\end{equation}
This (as well as a similar equation for an insertion of $\overline
T$) is the \em conformal Ward identity\em. We derived it assuming
that the quantum theory could be defined by a path integral and
the change in the action $\delta S$ follows the classical pattern.
For a more general CFT, not necessarily `defined' (however
loosely) by a path integral, (\ref{WI}) is usually assumed as a
property of $T$. In fact many basic introductions to CFT use this
as a starting point.

Scaling fields $\phi_j(z_j,\bar z_j)$ such that the most singular
term in their OPE with $T(z)$ is $O((z-z_j)^{-2})$ are called \em
primary\em.\footnote{If we assume that there is a lower bound to
the scaling dimensions, such fields must exist.} All the other
fields like those appearing in the less singular terms in
(\ref{Tope}) are called \em descendants\em. Once one knows the
correlation functions of all the primaries, those of the rest
follow from (\ref{Tope}).\footnote{Since the scaling dimensions of
the descendants differ from those of the corresponding primaries
by positive integers, they are increasingly irrelevant in the
sense of the renormalisation group.}

For correlations of such primary fields, we can now reverse the
arguments leading to (\ref{Tope}) for the case of a general
infinitesimal conformal transformation $\alpha(z)$ and conclude
that
$$
\delta\langle\phi_1(z_1,\bar z_1)\phi_2(z_2,\bar
z_2)\ldots\rangle=\sum_j(\Delta_j\alpha'(z_j)+\alpha(z_j)\partial_{z_j})
\langle\phi_1(z_1,\bar z_1)\phi_2(z_2,\bar z_2)\ldots\rangle\,,
$$
which may be integrated up to get the result for a \em finite \em
conformal mapping $z\to z'=f(z)$:
$$
\langle\phi_1(z_1,\bar z_1)\phi_2(z_2,\bar z_2)\ldots\rangle_{\cal
D}=\prod_jf'(z_j)^{\Delta_j}{\overline{f'(z_j)}}^{\bar \Delta_j}
\langle\phi_1(z'_1,{\bar z}'_1)\phi_2(z'_2,{\bar
z}'_2)\ldots\rangle_{{\cal D}'}\,.
$$
This is just the result we wanted to postulate in (\ref{conf}),
but now we see that it can hold only for correlation functions of
\em primary \em fields.

It is important to realise that $T$ itself is not in general
primary. Indeed its OPE with itself must take the
form\footnote{The $O((z-z_1)^{-3})$ term is absent by symmetry
under exchange of $z$ and $z_1$.}
\begin{equation}\label{TTope}
T(z)\cdot
T(z_1)=\frac{c/2}{(z-z_1)^4}+\frac2{(z-z_1)^2}T(z_1)+\frac1{z-z_1}\partial_{z_1}T(z_1)\cdots\,.
\end{equation}
This is because (taking the expectation value of both sides) the
2-point function $\langle T(z)T(z_1)\rangle$ is generally
non-zero. Its form is fixed by the fact that $\Delta_T=2$,
$\overline{\Delta}_T=0$, but, since the normalisation of $T$ is
fixed by its definition (\ref{dS}), its coefficient $c/2$ is
fixed. This introduces the \em conformal anomaly \em number $c$,
which is part of the basic data of the CFT, along with the scaling
dimensions $(\Delta_j,\overline{\Delta}_j)$  and the OPE
coefficients $c_{ijk}$.\footnote{When the theory is placed in
curved background, the trace $\langle T^\mu_\mu\rangle\propto cR$,
where $R$ is the local scalar curvature.}

This means that, under an infinitesimal transformation
$\alpha(z)$, there is an additional term in the transformation law
of $T$:
$$
\delta T(z)=2\alpha'(z)T(z)+\alpha(z)\partial_zT(z)+\ffrac
c{12}\alpha'''(z)\,.
$$
For a finite conformal transformation $z\to z'=f(z)$, this
integrates up to
\begin{equation}\label{Ttrans}
T(z)=f'(z)^2T(z')+\ffrac c{12}\{z',z\}\,,
\end{equation}
where the last term is the Schwarzian derivative
$$
\{w(z),z\}=\frac{w'''(z)w'(z)-\ffrac32w''(z)^2}{w'(z)^2}\,.
$$

The form of the Schwarzian can be seen most easily in the example
of a gaussian free field. In this case, the point-split terms in
(\ref{Tff}) transform properly and give rise to the first term in
(\ref{Ttrans}), but the fact that the subtraction has to be made
separately in the origin frame and the transformed one leads to an
anomalous term
$$
\lim_{\delta\to0}g\left(\frac{f'(z+\ffrac12\delta)f'(z-\ffrac12\delta)}
{2g\big(f(z+\ffrac12\delta)-f(z-\ffrac12\delta)\big)^2}-\frac1{2g\delta^2}\right)\,,
$$
which, after some algebra, gives the second term in (\ref{Ttrans})
with $c=1$.\footnote{This is a classic example of an anomaly in
QFT, which comes about because the regularisation procedure does
not respect a symmetry.}

\subsection{An application - entanglement entropy} Let's take a
break from the development of the general theory and discuss how
the conformal anomaly number $c$ arises in an interesting (and
topical) physical context. Suppose that we have a massless
relativistic quantum field theory in 1+1 dimensions, whose
imaginary time behaviour therefore corresponds to a euclidean CFT.
(There are many condensed matter systems whose large distance
behaviour is described by such a theory.) The system is at zero
temperature and therefore in its ground state $|0\rangle$,
corresponding to a density matrix $\rho=|0\rangle\langle0|$.
Suppose an observer A has access to just part of the system, for
example a segment of length $\ell$ inside the rest of the system,
of total length $L\gg\ell$, observed by B. The measurements of A
and B are entangled. It can be argued that a useful measure of the
degree of entanglement is the \em entropy \em $S_A=-{\rm
Tr}_A\,\rho_A\log\rho_A$, where $\rho_A={\rm Tr}_B\,\rho$ is the
reduced density matrix corresponding to A.

\begin{figure}
\centering
\includegraphics[width=8cm]{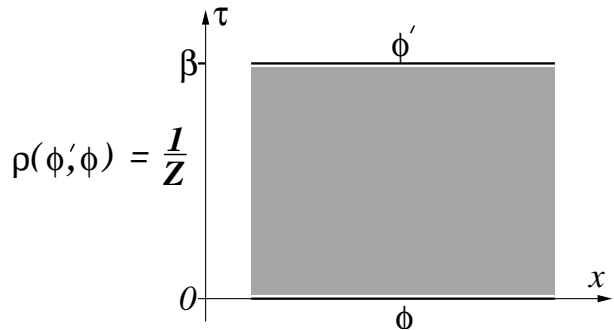}
\caption{\label{pathintegral}\small The density matrix is given by
the path integral over a space-time region in which the rows and
columns are labelled by the initial and final values of the
fields. }
\end{figure}

How can we calculate $S_A$ using CFT methods? The first step is to
realise that if we can compute ${\rm Tr}\,\rho_A^n$ for positive
integer $n$, and then analytically continue in $n$, the derivative
$\partial/\partial n|_{n=1}$ will give the required quantity. For
any QFT, the density matrix at finite inverse temperature $\beta$
is given by the path integral over some fundamental set of fields
$h(x,t)$ ($t$ is imaginary time)
$$
\rho\big(\{h(x,0)\},\{h(x,\beta)\}\big)=Z^{-1}\int'[dh(x,t)]e^{-S[h]}\,,
$$
where the rows and columns of $\rho$ are labelled by the values of
the fields at times $0$ and $\beta$ respectively, the the path
integral is over all histories $h(x,t)$ consistent with these
initial and final values (Fig.~\ref{pathintegral}).
$Z$ is the partition function, obtained by sewing together the
edges at these two times, that is setting $h(x,\beta)=h(x,0)$ and
integrating $\int[dh(x,0)]$.

\begin{figure}
\centering
\includegraphics[width=8cm]{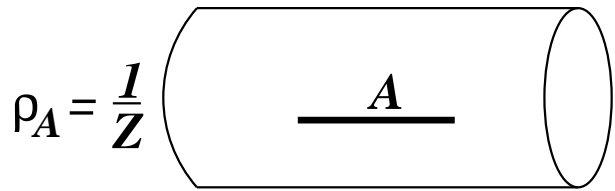}
\caption{\label{rhoa}\small The reduced density matrix $\rho_A$ is
given by the path integral over a cylinder with a slit along the
interval $A$.}
\end{figure}

\begin{figure}
\centering
\includegraphics[width=5cm]{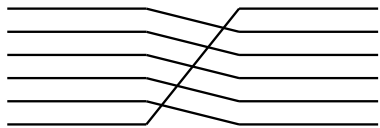}
\caption{\label{sewing}\small ${\rm Tr}\,\rho_A^n$ corresponds to
sewing together $n$ copies so that the edges are connected
cyclically.}
\end{figure}

We are interested in the partial density matrix $\rho_A$, which is
similarly found by sewing together the top and bottom edges for
$x\in B$, that is, leaving open a slit along the interval $A$
(Fig.~\ref{rhoa}).
The rows and columns of $\rho_A$ are labelled by the values of the
fields on the edges of the slit. ${\rm Tr}\,\rho_A^n$ is then
obtained by sewing together the edges of $n$ copies of this slit
cylinder in a cyclic fashion (Fig.~\ref{sewing}).
This gives an $n$-sheeted surface with branch points, or conical
singularities, at the ends of the interval $A$. If $Z_n$ is the
partition function on this surface, then
$$
{\rm Tr}\,\rho_A^n=Z_n/Z_1^n\,.
$$
Let us consider the case of zero temperature, $\beta\to\infty$,
when the whole system is in the ground state $|0\rangle$. If the
ends of the interval are at $(x_1,x_2)$, the conformal mapping
$$
z=\left(\frac{w-x_1}{w-x_2}\right)^{1/n}
$$
maps the $n$-sheeted $w$-surface to the single-sheeted complex
$z$-plane. We can use the transformation law (\ref{Ttrans}) to
compute $\langle T(w)\rangle$, given that $\langle T(z)\rangle=0$
by translational and rotational invariance. This gives, after a
little algebra,
$$
\langle T(w)\rangle=\frac{(c/12)(1-1/n^2)(x_2-x_1)^2}
{(w-x_1)^2(w-x_2)^2}\,.
$$
Now suppose we change the length $\ell=|x_2-x_1|$ slightly, by
making an infinitesimal non-conformal transformation $x\to
x+\delta\ell\delta(x-x_0)$, where $x_1<x_0<x_2$. The response of
the log of the partition function, by the definition of the stress
tensor, is
$$
\delta\log
Z_n=-\frac{n\delta\ell}{2\pi}\int_{-\infty}^\infty\langle
T_{xx}(x_0,t)\rangle dt
$$
(the factor $n$ occurs because it has to be inserted on each of
the $n$ sheets.) Writing $T_{xx}=T+\overline T$, the integration
in each term can be carried out by wrapping the contour around
either of the points $x_1$ or $x_2$. The result is
$$
\frac{\partial\log Z_n}{\partial\ell}=-\frac{(c/6)(n-1/n)}\ell\,,
$$
so that $Z_n/Z_1^n\sim\ell^{-(c/6)(n-1/n)}$. Taking the derivative
with respect to $n$ at $n=1$ we get the final result
$$
S_A\sim(c/3)\log\ell\,.
$$

\section{Radial quantisation and the Virasoro algebra}
\subsection{Radial quantisation}
\begin{figure}
\centering
\includegraphics[width=3cm]{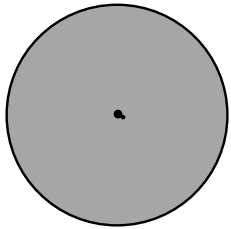}
\caption{\label{radialq}\small The state $|\phi_j\rangle$ is
defined by weighting field configurations on the circle with the
path integral inside, with an insertion of the field $\phi_j(0)$
at the origin.}
\end{figure}

Like any quantum theory CFT can be formulated in terms of local
operators acting on a Hilbert space of states. However, as it is
massless, the usual quantisation of the theory on the infinite
line is not so useful since it is hard to disentangle the
continuum of eigenstates of the hamiltonian, and we cannot define
asymptotic states in the usual way. Instead it is useful to
exploit the scale invariance, rather than the time-translation
invariance, and quantise on a circle of fixed radius $r_0$. In the
path integral formulation, heuristically the Hilbert space is the
space of field configurations $|\{h(r_0,\theta)\}\rangle$ on this
circle. The analogue of the hamiltonian is then the generator
$\hat D$ of scale transformations. It will turn out that the
spectrum of $\hat D$ is discrete. In the vacuum state each
configuration is weighted by the path integral over the disc
$|z|<r_0$, conditioned on taking the assigned values on $r_0$, see
Fig.~\ref{radialq}:
$$
|0\rangle=\int[dh(r\leq r_0)]e^{-S[h]}|\{h(r_0,\theta)\}\rangle\,.
$$
The scale invariance of the action means that this state is
independent of $r_0$, up to a constant.
If instead we insert a scaling field $\phi_j(0)$ into the above
path integral, we get a different state $|\phi_j\rangle$. On the
other hand, more general correlation functions of scaling fields
are given in this operator formalism by\footnote{We shall make an
effort consistently to denote actual operators (as opposed to
fields) with a hat.}
$$
\langle\phi_j(z_1,\bar z_1)\phi_j(z_2,\bar z_2)\rangle=
\langle0|{\bf R}\,\hat\phi_j(z_1,\bar z_1)\hat\phi_j(z_2,\bar
z_2)|0\rangle\,,
$$
where $\bf R$ means the $r$-ordered product (like the time-ordered
product in the usual case), and $\langle0|$ is defined similarly
in terms of the path integral over $r>r_0$. Thus we can identify
$$
|\phi_j\rangle=\lim_{z\to0}\hat\phi_j(z,\bar z)|0\rangle\,.
$$
This is an example of the \em operator-state correspondence \em in
CFT.

Just as the hamiltonian in ordinary QFT can be written as an
integral over the appropriate component of the stress tensor, we
can write
$$
\hat D=\hat L_0+\hat{\overline L}_0\equiv\frac1{2\pi i}\int_Cz\hat
T(z)dz+\mbox{c.c.}\,,
$$
where $C$ is any closed contour surrounding the origin. This
suggests that we define more generally
$$
\hat L_n\equiv\frac1{2\pi i}\int_Cz^{n+1}\hat T(z)dz\,,
$$
and similarly $\hat{\overline L}_n$.

If there are no operator insertions inside $C$ it can be shrunk to
zero for $n\geq-1$, thus
$$
\hat L_n|0\rangle=0\qquad\mbox{for $n\geq-1$}\,.
$$
On the other hand, if there is an operator $\phi_j$ inserted at
the origin, we see from the OPE (\ref{Tope}) that
$$
\hat L_0|\phi_j\rangle=\Delta_j|\phi_j\rangle\,.
$$
If $\phi_j$ is \em primary\em, we further see from the OPE that
$$
\hat L_n|\phi_j\rangle=0\qquad\mbox{for $n\geq1$}\,,
$$
while for $n\leq-1$ we get states corresponding to descendants of
$\phi_j$.
\subsection{The Virasoro algebra}
Now consider $\hat L_m\hat L_n$ acting on some arbitrary state. In
terms of correlation functions this involves the contour integrals
$$
\int_{C_2}\frac{dz_2}{2\pi i}z_2^{m+1} \int_{C_1}\frac{dz_1}{2\pi
i}z_1^{n+1}T(z_2)T(z_2)\,,
$$
where $C_2$ lies \em outside \em $C_1$, because of the $\bf
R$-ordering. If instead we consider the operators in the reverse
order, the contours will be reversed. However we can then always
distort them to restore them to their original positions, as long
as we leave a piece of the $z_2$ contour wrapped around $z_1$.
This can be evaluated using the OPE (\ref{TTope}) of $T$ with
itself:
\begin{eqnarray*}
&&\int_C\frac{dz_1}{2\pi i}z_1^{n+1}\oint\frac{dz_2}{2\pi
i}z_2^{m+1}\left(\frac2{(z_2-z_1)^2}T(z_1)+\frac1{z_2-z_1}
\partial_{z_1}T(z_1)+\frac{c/2}{)z_2-z_1)^4}\right)\\
&&=\int_C\frac{dz_1}{2\pi
i}z_1^{n+1}\left(2(m+1)z_1^mT(z_1)+z_1^{m+1}\partial_{z_1}T(z_1)+\ffrac
c{12}m(m^2-1)z_1^{m-2}\right)\,.
\end{eqnarray*}
The integrals can then be re-expressed in terms of the $\hat L_n$.
This gives the \em Virasoro algebra\em:
\begin{equation}\label{Vir}
[\hat L_m,\hat L_n]=(m-n)\hat L_{m+n}+\ffrac
c{12}m(m^2-1)\delta_{m+n,0}\,,
\end{equation}
with an identical algebra satisfied by the $\hat{\overline L}_n$.
It should be stressed that (\ref{Vir}) is completely equivalent to
the OPE (\ref{TTope}), but of course algebraic methods are often
more efficient in understanding the structure of QFT. The first
term on the right hand side could have been foreseen if we think
of $\hat L_n$ as being the generator of infinitesimal conformal
transformations with $\alpha(z)\propto z^{n+1}$. Acting on
functions of $z$ this can therefore be represented by
$\hat\ell_n=z^{n+1}\partial_z$, and it is easy to check that these
satisfy (\ref{Vir}) without the central term (called the Witt
algebra.) However the $\hat L_n$ act on states of the CFT rather
than functions, which allows for the existence of the second term,
the central term. The form of this, apart from the undetermined
constant $c$, is in fact dictated by consistency with the Jacobi
identity. Note that there is a closed subalgebra generated by
$(\hat L_1,\hat L_0,\hat L_{-1})$, which corresponds to special
conformal transformations.

One consequence of (\ref{Vir}) is
$$
[\hat L_0,\hat L_{-n}]=n\hat L_{-n}\,,
$$
so that $\hat L_0(\hat L_{-n}|\phi_j\rangle)=(\Delta_j+n)\hat
L_{-n}|\phi_j\rangle$, which means that the $\hat L_n$ with $n<0$
act as raising operators for the weight, or scaling dimension,
$\Delta$, and those with $n>0$ act as lowering operators. The
state $|\phi_j\rangle$ corresponding to a primary operator is
annihilated by all the lowering operators. It is therefore a \em
lowest weight state\em.\footnote{In the literature this is often
called, confusingly, a highest weight state.} By acting with all
possible raising operators we build up a \em lowest weight
representation \em (called a Verma module) of the Virasoro
algebra:
\begin{eqnarray*}
&& \vdots\\
&&\hat L_{-3}|\phi_j\rangle, \hat L_{-2}\hat L_{-1}|\phi_j\rangle,
{\hat L_{-1}}^3|\phi_j\rangle;\\
&&\hat L_{-2}|\phi_j\rangle, {\hat L_{-1}}^2|\phi_j\rangle;\\
&&\hat L_{-1}|\phi_j\rangle;\\
&&|\phi_j\rangle\,.
\end{eqnarray*}

\subsection{Null states and the Kac formula}

One of the most important issues in CFT is whether, for a given
$c$ and $\Delta_j$, this representation is unitary, and whether it
is reducible (more generally, decomposable). It turns out that
these two are linked, as we shall see later. Decomposability
implies the existence of null states in the Verma module, that is,
some linear combination of states at a given level is itself a
lowest state. The simplest example occurs at level 2, if
$$
\hat L_n\left(\hat L_{-2}|\phi_j\rangle-(1/g){\hat
L_{-1}}^2|\phi_j\rangle\right)=0\,,
$$
for $n>0$ (the notation with $g$ is chosen to correspond to the
Coulomb gas later on.) By taking $n=1$ and $n=2$ and using the
Virasoro algebra and the fact that $|\phi_j\rangle$ is a lowest
weight state, we get
$$
\Delta_j=\frac{3g-2}{4}\,,\qquad c=\frac{(3g-2)(3-2g)}{g}\,.
$$
This is special case $(r,s)=(2,1)$ of the \em Kac formula\em: with
$c$ parametrised as above, if\footnote{Note that $g$ and $1/g$
given the same value of $c$, and that
$\Delta_{r,s}(g)=\Delta_{s,r}(1/g)$. This has led to endless
confusion in the literature.}
\begin{equation}\label{Kac}
\Delta_j=\Delta_{r,s}(g)\equiv\frac{(rg-s)^2-(g-1)^2}{4g}\,,
\end{equation}
then $|\phi_j\rangle$ has a null state at level $r\cdot s$. We
will not prove this here, but later will indicate how it is
derived from Coulomb gas methods.

Removing all the null states from a Verma module gives an
irreducible representation of the Virasoro algebra. Null states
(and all their descendants) can consistently be set to zero in a
given CFT. (This is no guarantee that they are in fact absent,
however.)

One important consequence of the null state is that the
correlation functions of $\phi_j(z,\bar z)$ satisfy linear
differential equations in $z$ (or $\bar z$) of order $rs$. The
case $rs=2$ will be discussed as an example in the last lecture.
This allows us in principle to calculate all the four-point
functions and hence the OPE coefficients.

\subsection{Fusion rules}
Let us consider the 3-point function
$$
\langle\phi_{2,1}(z_1)\phi_{r,s}(z_2)\phi_\Delta(z_3)\rangle\,,
$$
where the first two fields sit at the indicated places in the Kac
table, but the third is a general primary scaling field of
dimension $\Delta$. The form of this is given by (\ref{3pt}). If
we insert $\int_C (z-z_1)^{-1}T(z)dz$ into this correlation
function, where $C$ surrounds $z_1$ but not the other two points,
this projects out $L_{-2}\phi_{2,1}\propto
\partial_{z_1}^2\phi_{2,1}$. On the other hand, the full
expression is given by the Ward identity (\ref{WI}). After some
algebra, we find that this is consistent only if
$$
\Delta=\Delta_{r\pm1,s}\,,
$$
otherwise the 3-point function, and hence the OPE coefficient of
$\phi_\Delta$ in $\phi_{2,1}\cdot\phi_{r,s}$, vanishes.

This is an example of the \em fusion rules \em in action. It shows
that Kac operators compose very much like irreducible
representations of SU$(2)$, with the $r$ label playing the role of
spin $\frac12(r-1)$. The $s$-labels compose in the same way. More
generally the \em fusion rule coefficients \em $N_{ij}^k$ tell us
not only which OPEs can vanish, but which ones actually do appear
in a particular CFT.\footnote{They can actually take values
$\geq2$ if there are distinct primary fields with the same
dimension.} In this simplest case we have (suppressing the
$s$-indices for clarity)
$$
N_{rr'}^{r''}=\delta_{r'',r+r'-1}+\delta_{r'',r+r'-3}+\cdots+
\delta_{r'',|r-r'|+1}\,.
$$

A very important thing happens if $g$ is rational $=p/p'$. Then we
can write the Kac formula as
$$
\Delta_{r,s}=\frac{(rp-sp')^2-(p-p')^2}{4pp'}\,,
$$
and we see that $\Delta_{r,s}=\Delta_{p'-r,p-s}$, that is, the
same primary field sits at two different places in the rectangle
$1\leq r\leq p'-1,1\leq s\leq p-1$, called the Kac table. If we
now apply the fusion rules to these fields we see that we get
consistency between the different constraints only if the fusion
algebra is \em truncated\em, that fields within the rectangle do
not couple to those outside.

This shows that, at least at the level of fusion, we can have CFTs
with a \em finite number \em of primary fields. These are called
the \em minimal models\em\footnote{The minimal models are examples
of \em rational \em CFTs: those which have only a finite number of
fields which are primary with respect to some algebra, more
generally an extended one containing Virasoro as a sub-algebra.}.
However, it can be shown that, among these, the only ones
admitting \em unitary \em representations of the Virasoro algebra,
that is for which $\langle\psi|\psi\rangle\geq0$ for all states
$|\psi\rangle$ in the representation, are those with $|p-p'|=1$
and $p,p'\geq3$. Moreover these are the \em only \em unitary CFTs
with $c<1$. The physical significance of unitarity will be
mentioned shortly.

\section{CFT on the cylinder and torus}
\subsection{CFT on the cylinder}
One of the most important conformal mappings is the logarithmic
transformation $w=(L/2\pi)\log z$, which maps the $z$-plane
(punctured at the origin) to an infinitely long cylinder of
circumference $L$ (equivalently, a strip with periodic boundary
conditions.) It is useful to write $w=t+iu$, and to think of the
coordinate $t$ running along the cylinder as imaginary time, and
$u$ as space. CFT on the cylinder then corresponds to euclidean
QFT on a circle.

The relation between the stress tensor on the cylinder and in the
plane is given by (\ref{Ttrans}):
$$
T(w)_{\rm cyl}=(dz/dw)^2T(z)+\ffrac
c{12}\{z,w\}=\left(2\pi/L\right)^2\left(z^2T(z)_{\rm plane}-\ffrac
c{24}\right)\,,
$$
where the last term comes from the Schwarzian derivative.

The hamiltonian $\hat H$ on the cylinder, which generates
infinitesimal translations in $t$, can be written in the usual way
as an integral over the time-time component of the stress tensor
$$
\hat H=\frac1{2\pi}\int_0^L\hat
T_{tt}(u)du=\frac1{2\pi}\int_0^L\big(\hat T(u)+\hat{\overline
T}(u)\big)du\,,
$$
which corresponds in the plane to
\begin{equation}\label{HL}
\hat H= \frac{2\pi}L\left(\hat L_0+\hat{\overline
L}_0\right)-\frac{\pi c}{6L}\,.
\end{equation}
Similarly the total momentum $\hat P$, which generates
infinitesimal translations in $u$, is the integral of the $T_{t
u}$ component of the stress tensor, which can be written as
$(2\pi/L)(\hat L_0-\hat{\overline L}_0)$.

Eq.~(\ref{HL}), although elementary, is one of the most important
results of CFT in two dimensions. It relates the dimensions of all
the scaling fields in the theory (which, recall, are the
eigenvalues of $\hat L_0$ and $\hat{\overline L}_0$) to the
spectra of $\hat H$ and $\hat P$ on the cylinder. If we have a
lattice model on the cylinder whose scaling limit is described by
a given CFT, we can therefore read off the scaling dimensions, up
to finite-size corrections in $(a/L)$, by diagonalising the
transfer matrix $\hat t\approx 1-a\hat H$. This can be done either
numerically for small values of $L$, or, for integrable models, by
solving the Bethe ansatz equations.

In particular, we see that the lowest eigenvalue of $\hat H$
(corresponding to the largest eigenvalue of the transfer matrix)
is
$$
E_0=-\frac{\pi
c}{6L}+\frac{2\pi}L(\Delta_0+\overline{\Delta}_0)\,,
$$
where $(\Delta_0,\overline{\Delta}_0)$ are the lowest possible
scaling dimensions. In many CFTs, and all unitary ones, this
corresponds to the identity field, so that
$\Delta_0=\overline{\Delta}_0=0$. This shows that $c$ can be
measured from finite-size behaviour of the ground state energy.

$E_0$ also gives the leading term in the partition function
$Z={\rm Tr}\,e^{-\ell\hat H}$ on a finite cylinder (a torus) of
length $\ell\gg L$. Equivalently, the free energy (in units of
$k_BT$) is
$$
F=-\log Z\sim-\frac{\pi c\ell}{6L}\,.
$$
In this equation $F$ represents the scaling limit of free energy
of a 2d \em classical \em lattice model\footnote{This treatment
overlooks UV divergent terms in $F$ of order $(\ell L/a^2)$, which
are implicitly set to zero by the regularisation of the stress
tensor.}. However we can equally well think of $t$ as being space
and $u$ imaginary time, in which case periodic boundary conditions
imply finite inverse \em temperature \em $\beta=1/k_BT=L$ in a 1d
\em quantum \em field theory. For such a theory we then predict
that
$$
F\sim -\frac{\pi c\ell k_BT}6\,,
$$
or, equivalently, that the low-temperature specific heat, at a
quantum critical point described by a CFT (generally, with a
linear dispersion relation $\omega\sim|q|$), has the form
$$
C_v\sim\frac{\pi ck_B^2T}3\,.
$$

Note that the Virasoro generators can be written in terms of the
stress tensor on the cylinder as
$$
\hat L_n=\frac L{2\pi}\int_0^Le^{inu}\hat T(u,0)du\,.
$$
In a unitary theory, $\hat T$ is self-adjoint, and hence $\hat
L_n^{\dag}=\hat L_{-n}$. Unitarity of the QFT corresponds to \em
reflection positivity \em of correlation functions: in general
$$
\langle\phi_1(u_1,t_1)\phi_2(u_2,t_2)\ldots
\phi_1(u_1,-t_1)\phi_2(u_2,-t_2)\rangle
$$
is positive if the transfer matrix $\hat T$ can be made
self-adjoint, which is generally true if the Boltzmann weights are
positive. Note however, that a given lattice model (e.g. the Ising
model) contains fields which are the scaling limit of lattice
quantities in which the transfer matrix can be locally expressed
(e.g. the local magnetisation and energy density) and for which
one would expect reflection positivity to hold, and other scaling
fields (e.g. the probability that a given edge lies on a cluster
boundary) which are not locally expressible. Within such a CFT,
then, one would expect to find a unitary \em sector \em -- in fact
in the Ising model this corresponds to the $p=3,p'=4$ minimal
model -- but also possible non-unitary sectors in addition.

\subsection{Modular invariance on the torus}
We have seen that unitarity (for $c<1$), and, more generally,
rationality, fix which scaling fields may appear in a given CFT,
but they don't fix which ones actually appear. This is answered by
considering another physical requirement: that of modular
invariance on the torus.

\begin{figure}
\centering
\includegraphics[width=5cm]{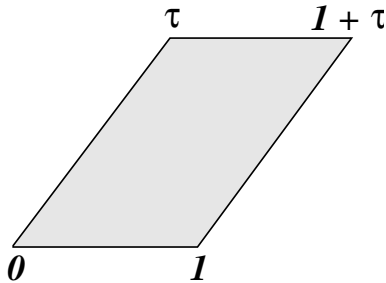}
\caption{\label{torus}\small A general torus is obtained by
identifying opposite sides of a parallelogram.}
\end{figure}

We can make a general torus by imposing periodic boundary
conditions on a parallelogram, whose vertices lie in the complex
plane. Scale invariance allows us to fix the length of one of the
sides to be unity: thus we can choose the vertices to be at
$(0,1,1+\tau,\tau)$, where $\tau$ is a complex number with
positive imaginary part. In terms of the conventions of the
previous section, we start with a finite cylinder of circumference
$L=1$ and length ${\rm Im}\,\tau$, twist one end by an amount
${\rm Re}\,\tau$, and sew the ends together -- see
Fig.~\ref{torus}.
\begin{figure}
\centering
\includegraphics[width=7cm]{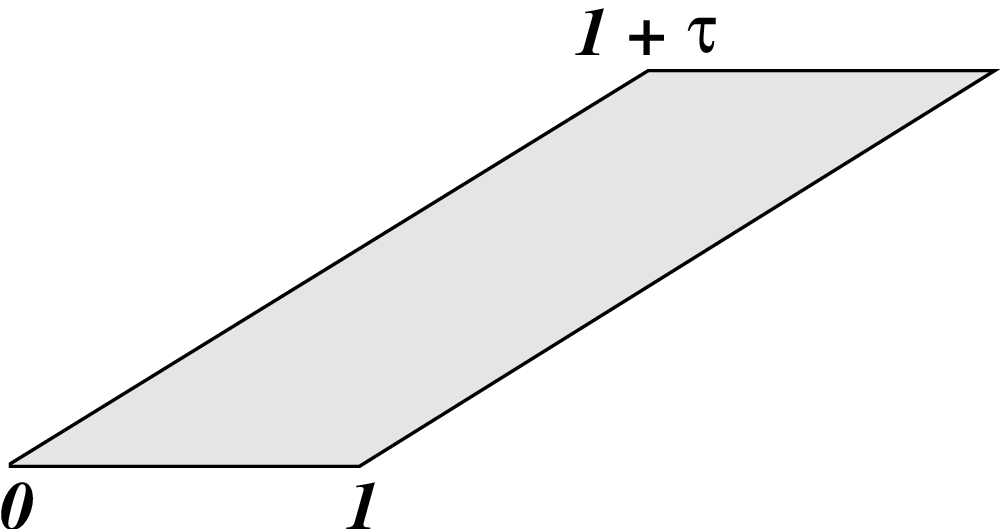}
\includegraphics[width=5cm]{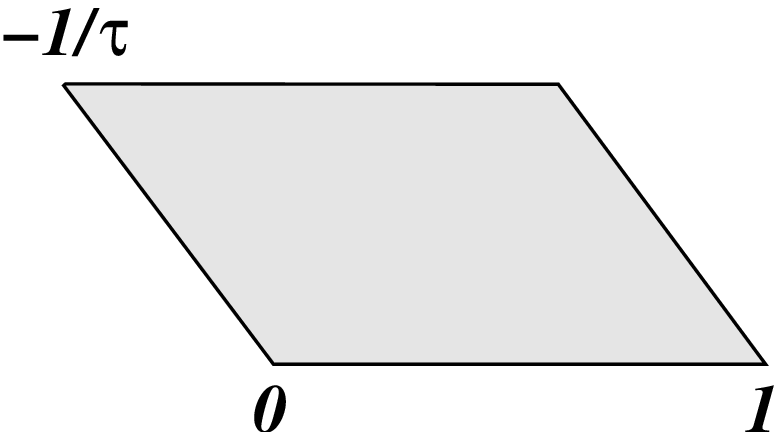}
\caption{\label{mod}\small Two ways of viewing the same torus,
corresponding to the modular transformations $T$ and $S$.}
\end{figure}
An important feature of this parametrisation of the torus is that
it is not unique: the transformations $T:\tau\to\tau+1$ and
$S:\tau\to-1/\tau$ give the same torus (Fig.~\ref{mod}). Note that
$S$ interchanges space $u$ and imaginary time $t$ in the QFT. $S$
and $T$ generate an infinite discrete group of transformations
$$
\tau\to\frac{a\tau+b}{c\tau+d}
$$
with $(a,b,c,d)$ all integers and $ad-bc=1$. This is called
SL$(2,{\bf Z})$ or the \em modular group\em. Note that $S^2=1$ and
$(ST)^3=1$.

Consider the scaling limit of the partition function $Z$ of a
lattice model on this torus. Apart from the divergent term in
$\log Z$, proportional to the area divided by $a^2$, which in CFT
is set to zero by regularisation, the rest should depend only on
the aspect ratio of the torus and thus be \em modular
invariant\em. This would be an empty statement were it not that
$Z$ can be expressed in terms of the spectrum of scaling
dimensions of the CFT in a manner which is not, manifestly,
modular invariant.

Recall that the generators of infinitesimal translations along and
around the cylinder can be written as
$$
\hat H=2\pi(\hat L_0+\hat{\overline L}_0)-\ffrac{\pi c}6\qquad
\hat P =2\pi(\hat L_0-\hat{\overline L}_0)\,.
$$
The action of twisting the cylinder corresponds to a finite
translation around its circumference, and sewing the ends together
corresponds to taking the trace. Thus
\begin{eqnarray*}
Z&=&{\rm Tr}\,e^{-({\rm Im}\,\tau)\hat H+i({\rm Re}\tau)\hat P}\\
&=&e^{\pi c{\rm Im}\,\tau/6}\,{\rm Tr}\,e^{2\pi i\tau\hat L_0}\,
e^{-2\pi i\hat{\overline L}_0}\\
&=&(q\bar q)^{-c/24}\,{\rm Tr}\,q^{\hat L_0}\,{\bar
q}^{\hat{\overline L}_0}\,,
\end{eqnarray*}
where in the last line we have defined $q\equiv e^{2\pi i\tau}$.

The trace means that we sum over all eigenvalues of $\hat L_0$ and
$\hat{\overline L}_0$, that is all scaling fields of the CFT. We
know that these can be organised into irreducible representations
of the Virasoro algebra, and therefore have the form
$(\Delta+N,\overline{\Delta}+\bar N)$, where $\Delta$ and
$\overline{\Delta}$ correspond to primary fields and $(N,\bar N)$
are non-negative integers labelling the levels of the descendants.
Thus we can write
$$
Z=\sum_{\Delta,\overline\Delta}n_{\Delta,\overline{\Delta}}\chi_\Delta(q)\chi_{\overline{\Delta}}(\bar
q)\,,
$$
where $n_{\Delta,\overline{\Delta}}$ is the number of primary
fields with lowest weights $(\Delta,\overline{\Delta})$, and
$$
\chi_\Delta(q)=q^{-c/24+\Delta}\sum_{N=0}^\infty d_\Delta(N)q^N\,,
$$
where $d_\Delta(N)$ is the degeneracy of the representation at
level $N$. It is purely a property of the representation, not the
particular CFT, and therefore so is $\chi_\Delta(q)$. This is
called the \em character \em of the representation.

The requirement of modular invariance of $Z$ under $T$ is rather
trivial: it says that all fields must have integer conformal
spin\footnote{It is interesting to impose other kinds of boundary
conditions, e.g. antiperiodic, on the torus, when other values of
the spin can occur.}. However the invariance under $S$ is highly
non-trivial: it states that $Z$, which is a power series in $q$
and $\bar q$, can equally well be expressed as an identical power
series in $\tilde q\equiv e^{-2\pi i/\tau}$ and $\bar{\tilde q}$.

We can get some idea of the power of this requirement by
considering the limit $q\to1$, $\tilde q\to0$, with $q$ real.
Suppose the density of scaling fields (including descendants) with
dimension $x=\Delta+\overline{\Delta}$ in the range $(x,x+\delta
x)$ (where $1\gg\delta x\gg x$) is $\rho(x)\delta x$. Then, in
this limit, when $q=1-\epsilon$, $\epsilon\ll1$,
$$
Z\sim\int^\infty \rho(x)e^{x\log
q}dx\sim\int^\infty\rho(x)e^{-\epsilon x}dx\,.
$$
On the other hand, we know that as $\tilde q\to0$, $Z\sim {\tilde
q}^{-c/12+x_0}\sim e^{(2\pi)^2(c/12-x_0)/\epsilon}$ where $x_0\leq
0$ is the lowest scaling dimension (usually 0). Taking the inverse
Laplace transform,
$$
\rho(x)\sim\int e^{\epsilon
x+(2\pi)^2(c/12-x_0)/\epsilon}\frac{d\epsilon}{2\pi i}\,.
$$
Using the method of steepest descents we then see that, as
$x\to\infty$,
$$
\rho(x)\sim\exp\left(4\pi(c/12-x_0)^{1/2}\,x^{1/2}\right)\,,
$$
times a (calculable) prefactor. This relation is of importance in
understanding black hole entropy in string theory.

\subsubsection{Modular invariance for the minimal models}
Let us apply this to the minimal models, where there is a finite
number of primary fields, labelled by $(r,s)$. We need the
characters $\chi_{r,s}(q)$. If there were no null states, the
degeneracy at level $N$ would be the number of states of the form
$\ldots \hat L_{-3}^{n_3}\hat L_{-2}^{n_2}\hat
L_{-1}^{n_1}|\phi\rangle$ with $\sum_jjn_j=N$. This is just the
number of distinct partitions of $N$ into positive integers, and
the generating function is $\prod_{k=1}^\infty(1-q^k)^{-1}$.

However, we know that the representation has a null state at level
$rs$, and this, and all its descendants, should be subtracted off.
Thus
$$
\chi_{rs}(q)=q^{-c/24}\prod_{k=1}^\infty(1-q^k)^{-1}\big(1-q^{rs}+\cdots\big)\,.
$$
But, as can be seen from the Kac formula (\ref{Kac}),
$\Delta_{r,s}+rs=\Delta_{p'+r,p-s}$, and therefore the null state
at level $rs$ has itself null states in its Verma module, which
should not have been subtracted off. Thus we must add these back
in. However, it is slightly more complicated than this, because
for a minimal model each primary field sits in two places in the
Kac rectangle, $\Delta_{r,s}=\Delta_{p'-r,p-s}$. Therefore this
primary field also has a null state at level $(p'-r)(p-s)$, and
this has the dimension $\Delta_{2p'-r,s}=\Delta_{r,2p-s}$ and
should therefore also be added back in if it has not be included
already. A full analysis requires understanding how the various
submodules sit inside each other, but fortunately the final result
has a nice form
\begin{equation}\label{char}
\chi_{rs}(q)=q^{-c/24}\prod_{k=1}^\infty(1-q^k)^{-1}\left(K_{r,s}(q)-K_{r,-s}(q)\right)\,,
\end{equation}
where
\begin{equation}\label{K}
K_{r,s}(q)=\sum_{n=-\infty}^\infty q^{(2npp'+rp-sp')^2/4pp'}\,.
\end{equation}

The partition function can then be written as finite sum
$$
Z=\sum_{r,s;\bar r,\bar s}n_{r,s;\bar r,\bar
s}\chi_{rs}(q)\chi_{\bar r\bar s}(\bar q)=\sum_{r,s;\bar r,\bar
s}n_{r,s;\bar r,\bar s}\chi_{rs}(\tilde q)\chi_{\bar r\bar
s}(\bar{\tilde q})\,.
$$
The reason this can happen is that the characters themselves
transform linearly under $S:q\to\tilde q$, as can be seen (after
quite a bit of algebra, by applying the Poisson sum formula to
(\ref{K}) and Euler's identities to the infinite product):
$$
\chi_{rs}(\tilde q)=\sum_{r',s'}S^{r's'}_{rs}\chi_{r's'}(q)\,,
$$
where $\bf S$ is a matrix whose rows and columns are labelled by
$(rs)$ and $(r's')$. Another way to state this is that the
characters form a representation of the modular group. The form of
$\bf S$ is not that important, but we give it anyway:
$$
S^{r's'}_{rs}=\left(\frac8{pp'}\right)^{1/2}(-1)^{1+r\bar s+s\bar
r} \sin\frac{\pi p'r\bar r}{p'}\sin\frac{\pi p's\bar s}{p}\,.
$$
The important properties of $\bf S$ are that it is real and
symmetric and ${\bf S}^2=1$.

This immediately implies that the \em diagonal \em combination,
with $n_{r,s;\bar r,\bar s}=\delta_{r\bar r}\delta_{s\bar s}$, is
modular invariant:
$$
\sum_{r,s}\chi_{rs}(\tilde q)\chi_{rs}(\bar{\tilde q})=
\sum_{r,s}\sum_{r',s'}\sum_{r'',s''}S^{r's'}_{rs}S^{r''s''}_{rs}
\chi_{r's'}(q)\chi_{r''s''}(\bar
q)=\sum_{r,s}\chi_{rs}(q)\chi_{rs}(\bar q)\,,
$$
where we have used ${\bf S}{\bf S}^T=1$. This gives the diagonal
series of CFTs, in which all possible scalar primary fields in the
Kac rectangle occur just once. These are known as the A$_n$
series.

It is possible to find other modular invariants by exploiting
symmetries of $\bf S$. For example, if $p'/2$ is odd, the space
spanned by $\chi_{rs}+\chi_{p'-r,s}$, with $r$ odd, is an
invariant subspace of $\bf S$, and is multiplied only by a pure
phase under $T$. Hence the diagonal combination within this
subspace
$$
Z=\sum_{r\ {\rm odd},s}|\chi_{r,s}+\chi_{p'-r,s}|^2
$$
is modular invariant. Similar invariants can be constructed if
$p/2$ is odd. This gives the D$_n$ series. Note that in this case
some fields appear with degeneracy 2. Apart from these two
infinite series, there are three special values of $p$ and $p'$
(12,18,30) denoted by E$_{6,7,8}$. The reason for this
classification will become more obvious in the next section when
we construct explicit lattice models which have these CFTs as
their scaling limits.\footnote{This classification also arises in
the finite subgroups of SU$(2)$, of simply-laced Lie algebras, and
in catastrophe theory.}

\section{Height models, loop models and Coulomb gas methods}
\subsection{Height models and loop models}
Although the ADE classification of minimal CFTs with $c<1$ through
modular invariance was a great step forwards, one can ask whether
there are in fact lattice models which have these CFTs as their
scaling limit. The answer is yes -- in the form of the ADE lattice
models. These can be analysed non-rigorously by so-called Coulomb
gas methods.

\begin{figure}[h]
\centering
\includegraphics[width=5cm]{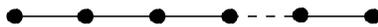}
\caption{\label{an}\small The graph $A_m$, with $m$ nodes.}
\end{figure}

For simplicity we shall describe only the so-called dilute models,
defined on a triangular lattice.\footnote{Similar models can be
defined on the square lattice. They give rise to critical loop
models in the dense phase.} At each site $r$ of the lattice is
defined a `height' $h(r)$ which takes values on the nodes of some
connected graph $\cal G$. An example is the linear graph called
$A_{m}$ shown in Fig.~\ref{an}, in which $h(r)$ can be thought of
as an integer between 1 and $m$.
There is a restriction in these models that the heights on
neighbouring sites of the triangular lattice must either be the
same, or be adjacent on $\cal G$. It is then easy to see that
around a given triangle either all three heights are the same
(which carries relative weight 1), or two of them are the same and
the other is adjacent on $\cal G$.\footnote{Apart from the
pathological case when $\cal G$ itself has a 3-cycle, in which
case we can enforce the restriction by hand.} In this case, if the
heights are $(h,h',h')$, the weight is $x(S_h/S_{h'})^{1/6}$ where
$S_h$ is a function of the height $h$, to be made explicit later,
and $x$ is a positive temperature-like parameter.\footnote{It is
sometimes useful, e.g. for implementing a transfer matrix, to
redistribute these weights around the loops.} (A simple example is
$A_2$, corresponding to the Ising model on the triangular
lattice.)

The weight for a given configuration of the whole lattice is the
product of the weights for each elementary triangle. Note that
this model is local and has positive weights if $S_h$ is a
positive function of $h$. Its scaling limit at the critical point
should correspond to a unitary CFT.

\begin{figure}
\centering
\includegraphics[width=5cm]{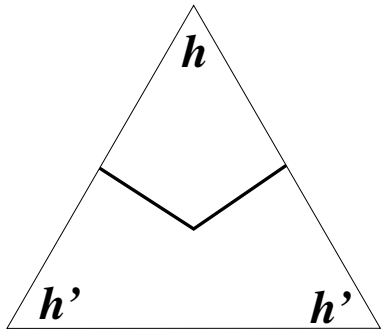}
\caption{\label{triangle}\small If the heights on the vertices are
not all equal, we denote this by a segment of a curve on the dual
lattice, as shown.}
\end{figure}

The height model can be mapped to a loop model as follows: every
time the heights in a given triangle are not all equal, we draw a
segment of a curve through it, as shown in Fig.~\ref{triangle}.
These segments all link up, and if we demand that all the heights
on the boundary are the same, they form a set of nested,
non-intersecting closed loops on the dual honeycomb lattice,
separating regions of constant height on the original lattice.
Consider a loop for which the heights just inside and outside are
$h$ and $h'$ respectively. The loop has convex (outward-pointing)
and concave (inward-pointing) corners. Each convex corner carries
a factor $(S_h/S_{h'})^{1/6}$, and each concave corner the inverse
factor. But each loop has exactly 6 more outward pointing corners
than inward pointing ones, so it always carries an overall weight
$S_h/S_h'$, times a factor $x$ raised to the length of the loop.
Let us now sum over the heights consistent with a fixed loop
configuration, starting with innermost regions. Each sum has the
form
$$
\sum_{h:|h-h'|=1}(S_h/S_{h'})\,,
$$
where $|h-h'|=1$ means that $h$ and $h'$ are adjacent on $\cal G$.
The next stage in the summation will be simple only if this is a
constant independent of $h'$. Thus we assume that the $S_h$
satisfy
$$
\sum_{h:|h-h'|=1}S_h=\Lambda S_{h'}\,,
$$
that is, $S_h$ is an eigenvector of the \em adjacency matrix \em
of $\cal G$, with eigenvalue $\Lambda$. For $A_{m}$, for example,
these have the form
\begin{equation}\label{eigen}
S_h\propto \sin\frac{\pi k h}{m+1}\,,
\end{equation}
where $1\leq k\leq m$, corresponding to $\Lambda=2\cos\big(\pi
k/(m+1)\big)$. Note that only the case $k=1$ gives all real
positive weights.

\begin{figure}
\centering
\includegraphics[width=8cm]{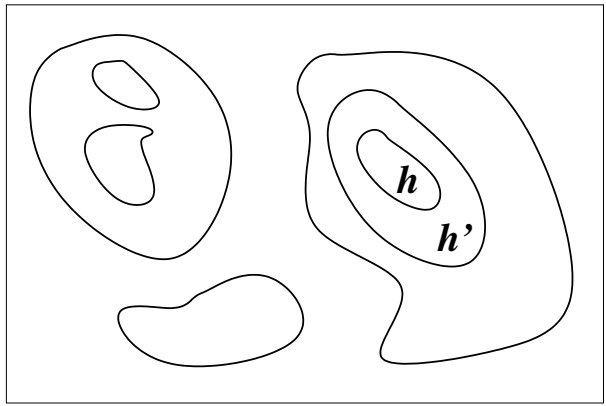}
\caption{\label{loops}\small Nested set of loops, separating
regions of constant height. We iteratively sum over the heights
$h$, starting with the innermost.}
\end{figure}
Having chosen the $S_h$ in this way, we can sum out all the
heights consistent with a given loop configuration
(Fig.~\ref{loops}), starting with the innermost and moving
outwards, and thereby express the partition function as a sum over
loop configurations:
\begin{equation}\label{Zloop}
Z=\sum_{\rm loop\ configs} \Lambda^{\rm number\ of\ loops}\,x^{\rm
total\ length}\,.
\end{equation}
When $x$ is small, the heights are nearly all equal (depending on
the boundary condition,) and the typical loop length and number is
small. At a critical point $x=x_c$ we expect this to diverge.
Beyond this, we enter the \em dense phase\em, which is still
critical in the loop sense, even though observables which are
local in the original height variables may have a finite
correlation length. For example, for $x>x_c$ in the Ising model,
the Ising spins are disordered but the the cluster boundaries are
the same, in the scaling limit, as those of critical percolation
for site percolation on the triangular lattice.

However, we could have obtained the same expression for $Z$ in
several different ways. One is by introducing $n$-component spins
$s_a(R)$ with $a=1,\ldots,n$ on the sites of the dual lattice, and
the partition function
$$
Z_{O(n)}={\rm
Tr}\,\prod_{RR'}\left(1+x\sum_{a=1}^ns_a(R)s_a(R')\right)\,,
$$
where the product is over edges of the honeycomb lattice, the
trace of an odd power of $s_a(R)$ is zero, and ${\rm
Tr}s_a(R)s_b(R)=\delta_{ab}$. Expanding in powers of $x$, and
drawing in a curve segment each time the term proportional to $x$
is chosen on a given edge, we get the same set of nested loop
configurations, weighted as above with $\Lambda=n$. This is the
O$(n)$ model. Note that the final expression makes sense for all
real positive values of $n$, but it can be expressed in terms of
weights local in the original spins only for positive integer $n$.
Only in the latter case do we therefore expect that correlations
of the O$(n)$ spins will satisfy reflection positivity and
therefore correspond to a unitary CFT, even though the description
in terms of heights is unitary. This shows how different sectors
of the `same' CFT can describe rather different physics.
\subsection{Coulomb gas methods}
These loop models can be solved in various ways, for example by
realising that their transfer matrix gives a representation of the
Temperley-Lieb algebra, but a more powerful if less rigorous
method is to use the so-called Coulomb gas approach. Recalling the
arguments of the previous section, we see that yet another way of
getting to (\ref{Zloop}) is by starting from a different height
model, where now the heights $h(r)$ are defined on the integers
(times $\pi$, for historical reasons) $\pi{\bf Z}$. As long as we
choose the correct eigenvalue of the adjacency matrix, this will
give the same loop gas weights. That is we take $\cal G$ to be
$A_\infty$. In this case the eigenvectors correspond to plane wave
modes propagating along the graph, labelled by a quasi-momentum
$\chi$ with $|\chi|<1$: $S_h\propto e^{i\chi h}$, whose eigenvalue
is $\Lambda=2\cos(\pi\chi)$. Because these modes are chiral, we
have to orient the loops to distinguish between $\chi$ and
$-\chi$. Each oriented loop then gets weighted with a factor
$e^{\pm i\pi\chi/6}$ at each vertex of the honeycomb lattice it
goes through, depending on whether it turns to the left or the
right.

This version of the model, where the heights are unbounded, is
much easier to analyse, at least non-rigorously. In particular, we
might expect that in the scaling limit, after coarse-graining, we
can treat $h(r)$ as taking all real values, and write down an
effective field theory. This should have the property that it is
local, invariant under $h(r)\to h(r)+$ constant, and with no terms
irrelevant under the RG (that is, entering the effective action
with positive powers of $a$.) The only possibility is a free
gaussian field theory, with action
$$
S=(g_0/4\pi)\int(\nabla h)^2d^2r\,.
$$

However, this cannot be the full answer, because we know this
corresponds to a CFT with $c=1$. The resolution of this is most
easily understood by considering the theory on a long cylinder of
length $\ell$ and circumference $L\ll\ell$. Non-contractible loops
which go around the cylinder have the same number of inside and
outside corners, so they are incorrectly counted. This can be
corrected by inserting a factor $\prod_te^{i\chi h(t,0)}e^{-i\chi
h(t+1,0)}$, which counts each loop passing between $(t,0)$ and
$(t+1,0)$ with just the right factors $e^{\pm i\pi\chi}$. These
factors accumulate to $e^{i\chi h(-\ell/2,0)}e^{-i\chi
h(\ell/2,0)}$, corresponding to charges $\pm\chi$ at the ends of
the cylinder.
\begin{figure}
\centering
\includegraphics[width=5cm]{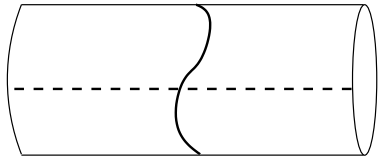}
\caption{\label{seam}\small Non-contractible loops on the cylinder
can be taken into account by the insertion of a suitable factor
along a seam.}
\end{figure}
This means that the partition function is
$$
Z\sim Z_{c=1}\langle e^{i\chi h(\ell/2)}e^{-i\chi
h(-\ell/2)}\rangle\,.
$$
But we know (Sec.~\ref{CWI}) that this correlation function decays
like $r^{-2x_{\chi}}$ in the plane, where $x_q=q^2/2g_0$, and
therefore on the cylinder
$$
Z\sim e^{\pi\ell/6L}\,e^{-2\pi(\chi^2/2g_0)\ell/L}\,,
$$
from which we see that the central charge is actually
$$
c=1-\frac{6\chi^2}{g_0}\,.
$$

However, we haven't yet determined $g_0$. This is fixed by the
requirement that the \em screening fields \em $e^{\pm i2 h(r)}$,
which come from the the fact that originally $h(r)\in\pi{\bf Z}$,
should be marginal, that is they do not affect the scaling
behaviour so that we can add them to the action with impunity.
This requires that they have scaling dimension $x_2=2$. However,
now $x_q$ should be calculated from the cylinder with the charges
$e^{\pm i\chi h(\pm\ell/2)}$ at the ends:
\begin{equation}\label{xq}
x_q=\frac{\big(q\pm\chi\big)^2}{g_0}-\frac{\chi^2}{g_0}
=\frac{q^2\pm2\chi q}{g_0}\,.
\end{equation}
Setting $x_2=2$ we then find $g_0=1\pm\chi$ and therefore
$$
c=1-\frac{6(g_0-1)^2}{g_0}\,.
$$

\subsection{Identification with minimal models}
The partition function for the height models (at least on
cylinder) depends only on the eigenvalue $\Lambda$ of the
adjacency matrix and hence the Coulomb gas should work equally
well for the models on $A_{m}$ if we set $\chi=k/(m+1)$. The
corresponding central charge is then
$$
c=1-\frac{6k^2}{(m+1)(m+1\pm k)}\,.
$$
If we compare this with the formula for the minimal models
$$
c=1-\frac{6(p-p')^2}{pp'}\,,
$$
we are tempted to identify $k=p-p'$ and $m+1=p'$. This implies
$g_0=p/p'$, which can therefore be identified with the parameter
$g$ introduced in the Kac formula.\footnote{The other solution
corresponds to interchanging $p'$ and $p$.} Moreover, if we
compute the scaling dimensions of local fields
$\phi_r(R)=\cos\big((r-1)kh(R)/(m+1)\big)$ using (\ref{xq}) we
find perfect agreement with the leading diagonal $\Delta_{r,r}$ of
the Kac table.\footnote{These are the relevant fields in the RG
sense.} We therefore have strong circumstantial evidence that the
scaling limit of the dilute $A_{p'-1}$ models (choosing the
eigenvalue $\Lambda=2\cos(\pi(p-p')/p')$) is the $(p,p')$ minimal
model with $p>p'$. Note that only if $k=1$, that is $p=p'+1$, are
these CFTs unitary, and this is precisely the case where the
weights of the lattice model are real and positive.

For other graphs $\cal G$ we can try to make a similar
identification. However, this is going to work only if the maximal
eigenvalue of the adjacency matrix of $\cal G$ is strictly less
than 2. A famous classification then shows that this restricts
$\cal G$ to be either of the form $A_m$, $D_m$ or one of three
exceptional cases $E_{6,7,8}$.\footnote{$\Lambda=2$ corresponds to
the extended diagrams $\hat A_m$, etc., which give interesting
rational CFTs with $c=1$. However models based on graphs with
$\Lambda>2$ probably have a different kind of a transition at
which the mean loop length remains finite.} These other graphs
also have eigenvalues of the same form (ref{eigen}) as $A_{m}$,
but with $m+1$ now being the Coxeter number, and the allowed
integers $k$ being only a subset of those appearing in the
$A$-series. These correspond to the Kac labels of the allowed
scalar operators which appear in the appropriate modular invariant
partition function.

\section{Boundary conformal field theory}
\subsection{Conformal boundary conditions and Ward identities}
So far we haven't considered what happens at the boundary of the
domain $\cal D$. This is a subject with several important
applications, for example to quantum impurity problems (see the
lectures by Affleck) and to D-branes in string theory.

In any field theory in a domain with a boundary, one needs to
consider how to impose a set of consistent boundary conditions.
Since CFT is formulated independently of a particular set of
fundamental fields and a lagrangian, this must be done in a more
general manner. A natural requirement is that the off-diagonal
component $T_{\parallel\perp}$ of the stress tensor
parallel/perpendicular to the boundary should vanish. This is
called the conformal boundary condition. If the boundary is
parallel to the time axis, it implies that there is no momentum
flow across the boundary. Moreover, it can be argued that, under
the RG, any uniform boundary condition will flow into a
conformally invariant one. For a given bulk CFT, however, there
may be many possible distinct such boundary conditions, and it is
one task of BCFT  to classify these.

To begin with, take the domain to be the upper half plane, so that
the boundary is the real axis. The conformal boundary condition
then implies that $T(z)=\overline T(\bar z)$ when $z$ is on the
real axis. This has the immediate consequence that correlators of
$\overline T$ are those of $T$ analytically continued into the
lower half plane. The conformal Ward identity now reads
\begin{eqnarray}
\langle T(z)\prod_j\phi_j(z_j,\bar z_j)\rangle
&=&\sum_j\left({\Delta_j\over(z-z_j)^2}+{1\over
z-z_j}\partial_{z_j}\right.
\nonumber\\
&&\qquad \left.+{\overline\Delta_j\over (z-\bar z_j)^2}+{1\over
z-\bar z_j}\partial_{\bar z_j}
\right)\langle\prod_j\phi_j(z_j,\bar z_j)\rangle\,. \label{BWI}
\end{eqnarray}

In radial quantisation, in order that the Hilbert spaces defined
on different hypersurfaces be equivalent, one must now choose
semicircles centered on some point on the boundary, conventionally
the origin. The dilatation operator is now
\begin{equation}
\hat D={1\over 2\pi i}\int_Sz\,\hat T(z)dz-{1\over 2\pi
i}\int_S\bar z\, \hat{\overline T}(\bar z)d\bar z\,,
\end{equation}
where $S$ is a semicircle. Using the conformal boundary condition,
this can also be written as
\begin{equation}
\label{DL} \hat D=\hat L_0={1\over 2\pi i}\int_Cz\,\hat T(z)dz\,,
\end{equation}
where $C$ is a complete circle around the origin.

Note that there is now only one Virasoro algebra. This is related
to the fact that conformal mappings which preserve the real axis
correspond to real analytic functions. The eigenstates of $\hat
L_0$ correspond to \em boundary operators \em $\hat \phi_j(0)$
acting on the vacuum state $|0\rangle$. It is well known that in a
renormalizable QFT fields at the boundary require a different
renormalization  from those in the bulk, and this will in general
lead to a different set of conformal weights. It is one of the
tasks of BCFT to determine these, for a given allowed boundary
condition.

However, there is one feature unique to boundary CFT in two
dimensions. Radial quantization also makes sense, leading to the
same form (\ref{DL}) for the dilation operator, if the boundary
conditions on the negative and positive real axes are \em
different\em. As far as the structure of BCFT goes, correlation
functions with this mixed boundary condition behave as though a
local scaling field were inserted at the origin. This has led to
the term `boundary condition changing (bcc) operator'.

\subsection{CFT on the annulus and classification of boundary
states} 
\begin{figure}
\centering
\includegraphics[width=5cm]{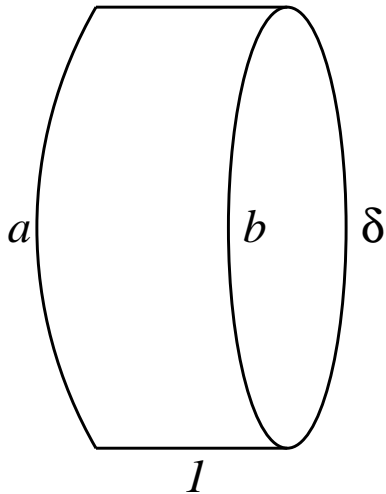}
\caption{\label{figannulus}\small The annulus, with boundary
conditions $a$ and $b$ on either boundary.}
\end{figure}

Just as consideration of the partition function on the
torus illuminates the bulk operator content
$n_{\Delta,\overline\Delta}$, it turns out that consistency on the
annulus helps classify both the allowed boundary conditions, and
the boundary operator content. To this end, consider a CFT in an
annulus formed of a rectangle of unit width and height $\delta$,
with the top and bottom edges identified (see
Fig.~\ref{figannulus}).
The boundary conditions on the left and right edges, labelled by
$a,b,\ldots$, may be different. The partition function with
boundary conditions $a$ and $b$ on either edge is denoted by
$Z_{ab}(\delta)$.

One way to compute this is by first considering the CFT on an
infinitely long strip of unit width. This is conformally related
to the upper half plane (with an insertion of boundary condition
changing operators at $0$ and $\infty$ if $a\not=b$) by the
mapping $z\to(1/\pi)\ln z$. The generator of infinitesimal
translations along the strip is
\begin{equation}
\hat H_{ab}=\pi\hat D-\pi c/24=\pi\hat L_0-\pi c/24\,.
\end{equation}
Thus for the annulus
\begin{equation}
Z_{ab}(\delta)={\rm Tr}\, e^{-\delta\,\hat H_{ab}} ={\rm
Tr}\,q^{\hat L_0-\pi c/24}\,,
\end{equation}
with $q\equiv e^{-\pi\delta}$. As before, this can be decomposed
into characters
\begin{equation}
\label{Zab}
Z_{ab}(\delta)=\sum_{\Delta}n_{ab}^{\Delta}\,\chi_{\Delta}(q)\,,
\end{equation}
but note that now the expression is linear. The non-negative
integers $n_{ab}^{\Delta}$ give the operator content with the
boundary conditions $(ab)$: the lowest value of ${\Delta}$ with
$n_{ab}^{\Delta}>0$ gives the conformal weight of the bcc
operator, and the others give conformal weights of the other
allowed primary fields which may also sit at this point.

On the other hand, the annulus partition function may be viewed,
up to an overall rescaling, as the path integral for a CFT on a
circle of unit circumference, being propagated for (imaginary)
time $\delta^{-1}$. From this point of view, the partition
function is no longer a trace, but rather the matrix element of
$e^{-\hat H/\delta}$ between \em boundary states\em:
\begin{equation}
\label{Zab2} Z_{ab}(\delta)=\langle a|e^{-\hat
H/\delta}|b\rangle\,.
\end{equation}
Note that $\hat H$ is the same hamiltonian that appears on the
cylinder, and the boundary states lie in the Hilbert space of
states on the circle. They can be decomposed into linear
combinations of states in the representation spaces of the two
Virasoro algebras, labelled by their lowest weights
$(\Delta,\overline{\Delta})$.

How are these boundary states to be characterized? Recalling that
on the cylinder $\hat L_n\propto\int e^{inu}\hat T(u)du$, and
$\hat{\overline L}_n\propto\int e^{-inu}\hat{\overline T}(u)du$,
the conformal boundary condition implies that any boundary state
$|B\rangle$ lies in the subspace satisfying
\begin{equation}
\label{LL} \hat L_n|B\rangle=\hat{\overline L}_{-n}|B\rangle\,.
\end{equation}
This condition can be applied in each subspace. Taking $n=0$ in
(\ref{LL}) constrains $\overline{\Delta}={\Delta}$. It can then be
shown that the solution of (\ref{LL}) is unique within each
subspace and has the following form. The subspace at level $N$ of
has dimension $d_{\Delta}(N)$. Denote an orthonormal basis by
$|{\Delta},N;j\rangle$, with $1\leq j\leq d_{\Delta}(N)$, and the
same basis for the representation space of $\overline{\rm Vir}$ by
$\overline{|{\Delta},N;j\rangle}$. The solution to (\ref{LL}) in
this subspace is then
\begin{equation}
|{\Delta}\rangle\rangle\equiv\sum_{N=0}^\infty\sum_{j=1}^{d_{\Delta}(N)}
|{\Delta},N;j\rangle\otimes\overline{|{\Delta},N;j\rangle}\,.
\end{equation}
These are called Ishibashi states. One way to understand this is
to note that (\ref{LL}) implies that
$$
\langle B|\hat L_n|B\rangle=\langle B|\hat{\overline
L}_{-n}|B\rangle=\langle B|hat{\overline L}_n|B\rangle\,,
$$
where we have used ${\hat{\overline L}}^{\dag}_{-n}=\hat{\overline
L}_n$ and assumed that the matrix elements are all real. This
means that acting with the raising operators $\hat{\overline L}_n$
on $|B\rangle$ has exactly the same effect as the $\hat L_n$, so,
starting with $N=0$ we must build up exactly the same state in the
two spaces.

 Matrix elements of the translation operator along the
cylinder between Ishibashi states are simple:
\begin{eqnarray}
&&\langle\langle {\Delta}'|e^{-\hat H/\delta}|{\Delta}\rangle\rangle\\
&&=\sum_{N'=0}^\infty\sum_{j'=1}^{d_{{\Delta}'}(N')}
\sum_{N=0}^\infty\sum_{j=1}^{d_{\Delta}(N)} \langle
{\Delta}',N';j'|\otimes\overline{\langle {\Delta}',N';j'|}
e^{-(2\pi/\delta)(\hat L_0+\hat{\overline L}_0-c/12)}\nonumber\\
&&\qquad\qquad\qquad\qquad\qquad\qquad\qquad\qquad\qquad
|{\Delta},N;j\rangle\otimes\overline{|{\Delta},N;j\rangle}\\
&&=\delta_{{\Delta}'{\Delta}}
\sum_{N=0}^\infty\sum_{j=1}^{d_{\Delta}(N)}e^{-(4\pi/\delta)\big({\Delta}+N-(c/24)\big)}
=\delta_{{\Delta}'{\Delta}}\,\chi_{\Delta}(e^{-4\pi/\delta})\,.
\end{eqnarray}
Note that the characters which appear are related to those in
(\ref{Zab}) by the modular transformation $S$.

The \em physical \em boundary states satisfying (\ref{Zab}) are
linear combinations of these Ishibashi states:
\begin{equation}
|a\rangle=\sum_{\Delta}\langle\langle
{\Delta}|a\rangle\,|{\Delta}\rangle\rangle\,.
\end{equation}
Equating the two different expressions (\ref{Zab},\ref{Zab2}) for
$Z_{ab}$, and using the modular transformation law for the
characters and their linear independence, gives the (equivalent)
conditions:
\begin{eqnarray}
n^{\Delta}_{ab}&=&\sum_{{\Delta}'}S_{{\Delta}'}^{\Delta}\langle
a|{\Delta}'\rangle\rangle
\langle\langle {\Delta}'|b\rangle\,;\label{C1}\\
\langle a|{\Delta}'\rangle\rangle \langle\langle
{\Delta}'|b\rangle
&=&\sum_{\Delta}S_{\Delta}^{{\Delta}'}n^{\Delta}_{ab}\,.\label{C2}
\end{eqnarray}
The requirements that the right hand side of (\ref{C1}) should
give a non-negative integer, and that the right hand side of
(\ref{C2}) should factorize in $a$ and $b$, give highly nontrivial
constraints on the allowed boundary states and their operator
content.

\begin{figure}
\centering
\includegraphics[width=9cm]{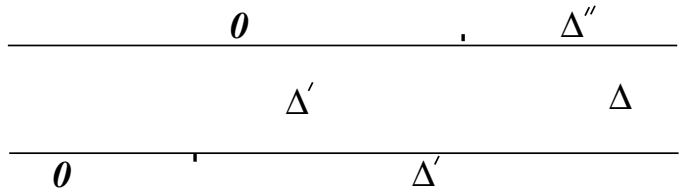}
\caption{\label{fusion}\small Argument illustrating the fusion
rules.}
\end{figure}

For the diagonal CFTs considered here (and for the nondiagonal
minimal models) a complete solution is possible. Since the
elements $S^{\Delta}_0$ of $\bf S$ are all non-negative, one may
choose $\langle\langle {\Delta}|\tilde
0\rangle=\big(S_0^{\Delta}\big)^{1/2}$. This defines a boundary
state
\begin{equation}
|\tilde
0\rangle\equiv\sum_{\Delta}\big(S_0^{\Delta}\big)^{1/2}|{\Delta}\rangle\rangle\,,
\end{equation}
and a corresponding boundary condition such that
$n^{\Delta}_{00}=\delta_{{\Delta}0}$. Then, for each
${\Delta}'\not=0$, one may define a boundary state
\begin{equation}
\langle\langle {\Delta}|\tilde{{\Delta}'}\rangle\equiv
S^{\Delta}_{{\Delta}'}/\big(S_0^{{\Delta}}\big)^{1/2}\,.
\end{equation}
From (\ref{C1}), this gives
$n^{\Delta}_{{\Delta}'0}=\delta_{{\Delta}'{\Delta}}$. For each
allowed ${\Delta}'$ in the torus partition function, there is
therefore a boundary state $|\tilde{{\Delta}'}\rangle$ satisfying
(\ref{C1},\ref{C2}). However, there is a further requirement:
\begin{equation}
\label{verl}
n^{\Delta}_{{\Delta}'{\Delta}''}=\sum_\ell{S^{\Delta}_\ell
S^\ell_{{\Delta}'}S^\ell_{{\Delta}''}\over S^\ell_0}
\end{equation}
should be a non-negative integer. Remarkably, this combination of
elements of $\bf S$ occurs in the \em Verlinde formula\em, which
follows from considering consistency of the CFT on the torus. This
states that the right hand side of (\ref{verl}) is equal to the
fusion rule coefficient $N^{\Delta}_{{\Delta}'{\Delta}''}$. Since
these are non-negative integers, the above ansatz for the boundary
states is consistent. The appearance of the fusion rules in this
context can be understood by the following argument, illustrated
in Fig.~\ref{fusion}.
Consider a very long strip. At `time' $t\to-\infty$ the boundary
conditions on both sides are those corresponding to $\tilde 0$, so
that only states in the representation $0$ propagate. At time
$t_1$ we insert the bcc operator $(0|\Delta')$ on one edge: the
states $\Delta'$ then propagate. This can be thought of as the
fusion of $0$ with $\Delta'$. At some much later time we insert
the bcc operator $(0|\Delta'')$ on the other edge: by the same
argument  this should correspond to the fusion of $\Delta'$ and
$\Delta''$, which gives all states $\Delta$ with
$N_{\Delta',\Delta''}^\Delta=1$. But by definition, these are
those with $n_{\Delta',\Delta''}^\Delta=1$.

We conclude that, at least for the diagonal models, there is a
bijection between the allowed primary fields in the bulk CFT and
the allowed conformally invariant boundary conditions. For the
minimal models, with a finite number of such primary fields, this
correspondence has been followed through explicitly.

\subsubsection{Example}
The simplest example is the diagonal $c=\frac12$ unitary CFT
corresponding to $p=4,p'=3$. The allowed values of the conformal
weights are $h=0,\frac12\frac1{16}$, and
\begin{equation}
{\bf S}=\pmatrix{\ffrac12&\ffrac12&\ffrac1{\sqrt2}\cr
                 \ffrac12&\ffrac12&-\ffrac1{\sqrt2}\cr
                 \ffrac1{\sqrt2}&-\ffrac1{\sqrt2}&0\cr}\,,
\end{equation}
from which one finds the allowed boundary states
\begin{eqnarray}
|\tilde0\rangle&=&\ffrac1{\sqrt2}|0\rangle\rangle
+\ffrac1{\sqrt2}|\ffrac12\rangle\rangle
+\ffrac1{2^{1/4}}|\ffrac1{16}\rangle\rangle\,;\\
|\tilde{\ffrac12}\rangle&=&\ffrac1{\sqrt2}|0\rangle\rangle
+\ffrac1{\sqrt2}|\ffrac12\rangle\rangle
-\ffrac1{2^{1/4}}|\ffrac1{16}\rangle\rangle\,;\\
|\tilde{\ffrac1{16}}\rangle&=&
|0\rangle\rangle-|\ffrac12\rangle\rangle\,.
\end{eqnarray}
The nonzero fusion rule coefficients  of this CFT are
$$
N_{0,0}^0=N_{0,\frac1{16}}^{\frac1{16}}=N_{0,\frac12}^{\frac12}=
N_{\frac1{16},\frac1{16}}^0=N_{\frac1{16},\frac1{16}}^{\frac12}
=N_{\frac12,\frac12}^0=N_{\frac1{16},\frac12}^{\frac1{16}}=1\,.
$$

The $c=\frac12$ CFT is known to describe the continuum limit of
the critical Ising model, in which spins $s=\pm1$ are localized on
the sites of a regular lattice. The above boundary conditions may
be interpreted as the continuum limit of the lattice boundary
conditions $s=1$, $s=-1$ and free ($f$), respectively. Note there
is a symmetry of the fusion rules which means that one could
equally well have reversed the first two. This shows, for example,
that the for $(ff)$ boundary conditions the states with lowest
weights $0$ (corresponding to the identity operator) and $\frac12$
(corresponding to the the magnetisation operator at the boundary)
can propagate. Similarly, the scaling dimension of the $(f|\pm1)$
bcc operator is $\frac1{16}$.

\subsection{Boundary operators and SLE} Let us now apply the above
ideas to the $A_{m}$ models. There should be a set of conformal
boundary states corresponding to the entries of first row $(r,1)$
of the Kac table, with $1\leq r\leq m$. It is an educated guess
(confirmed by exact calculations) that these in fact correspond to
lattice boundary conditions where the heights on the boundary are
fixed to be at a particular node $r$ of the $A_{m}$ graph. What
about the boundary condition changing operators? These are given
by the fusion rules. In particular, since (suppressing the index
$s=1$)
$$
N_{r,2}^{r'}=\delta_{|r-r'|,1}\,,
$$
we see that the bcc operator between $r$ and $r\pm1$,
corresponding to a single cluster boundary intersecting the
boundary of the domain, must be a $(2,1)$ operator in the Kac
table.\footnote{If instead of the dilute lattice model we consider
the dense phase, which corresponds, eg to the boundaries of the FK
clusters in the Potts model, then $r$ and $s$ get interchanged for
a given central charge $c$, and the bcc operator then lies at
$(1,2)$.} This makes complete sense: if we want to go from $r_1$
to $r_2$ we must bring together at least $|r_1-r_2|$ cluster
boundaries, showing that the leading bcc operator in this case is
at $(|r_1-r_2|,1)$, consistent once again with the fusion rules.
If the bcc operators corresponding to a single curve are $(2,1)$
this means that the corresponding states satisfy
\begin{equation}\label{NS}
\big(2\hat L_{-2}-(2/g){\hat L}_{-1}^2\big)|\phi_{2,1}\rangle=0\,.
\end{equation}

We are now going to argue that (\ref{NS}) is equivalent to the
statement that the cluster boundary starting at this boundary
point is described by SLE. In order to avoid being too abstract
initially, we'll first show how the calculations of a particular
observable agree in the two different formalisms.

Let $\zeta$ be a point in the upper half plane and let $P(\zeta)$
be the probability that the curve, starting at the origin, passes
to the left of this point (of course it is not holomorphic). First
we'll give the physicist's version of the SLE argument (assuming
familiarity with Werner's lectures). We imagine making the
exploration process for a small Loewner time $\delta t$, then
continuing the process to infinity. Under the conformal mapping
$f_{\delta t}(z)$ which removes the first part of the curve, we
get a new curve with the same measure as the original one, but the
point $\zeta$ is mapped to $f_{\delta t}(\zeta)$. But this will
lie to the right of the new curve if and only if the original
point lay to the right of the original curve. Also, by integrating
the Loewner equation starting from $f_0(z)=z$, we have
approximately
$$
f_{\delta t}(z)\approx z+\frac{2\delta t}{z}+\sqrt\kappa \delta
B_t\,,
$$
at least for $z\gg\delta t$. Thus we can write down an
equation\footnote{Some physicists will recognise this as the
reverse Fokker-Planck equation.}:
$$
P(\zeta)={\bf E}\left[P\left(\zeta+\frac{2\delta
t}\zeta+\sqrt\kappa \delta B_{t}\right)\right]_{\delta B_{t}}\,,
$$
where ${\bf E}[\ldots]_{\delta B_{t}}$ means an average over all
realisations of the Brownian motion up to time $\delta t$.
Expanding the right hand side to $O(\delta t)$, and remembering
that ${\bf E}[\delta B_{t}]=0$ and ${\bf E}[(\delta
B_{t})^2]=\delta t$, we find (with $\zeta=x+iy$), the linear PDE
\begin{equation}\label{pde}
\left(\frac{2x}{x^2+y^2}\frac{\partial}{\partial x} -
\frac{2y}{x^2+y^2}\frac{\partial}{\partial y}+\frac\kappa
2\frac{\partial^2}{\partial x^2}\right)P(x,y)=0\,.
\end{equation}
By scale invariance,  $P(x,y)$ depends in fact only on the ratio
$y/x$, and therefore this can be reduced to a second order ODE,
whose solution, with appropriate boundary conditions, can be
expressed in terms of hypergeometric functions (and is known as
Schramm's formula.)

Now let us give the CFT derivation. In terms of correlation
functions, $P$ can be expressed as
$$
P=\frac{\langle\phi_{2,1}(0)\Phi(\zeta,\bar\zeta)\phi_{2,1}(\infty)\rangle}
{\langle\phi_{2,1}(0)\phi_{2,1}(\infty)\rangle}\,.
$$
The denominator is just the partition function restricted to there
being a cluster boundary from $0$ to infinity. $\Phi$ is an
`indicator operator' which takes the values $0$ or $1$ depending
on whether the curve passes to the right (respectively left) of
$\zeta$. Since $P$ is a probability it is dimensionless, so $\Phi$
has zero conformal dimensions and transforms trivially.

Now suppose we insert $\int_C(2T(z)/z)(dz/2\pi i)+{\rm cc.}$ into
the correlation function in the numerator, where $C$ is a
semicircular contour surrounding the origin but not $\zeta$. Using
the OPE of $T$ with $\phi_{2,1}$, this gives
$$
2L_{-2}\phi_{2,1}(0)=(2/g)\partial_x^2\phi_{1,2}(x)|_{x=0}\,.
$$
Using translation invariance, this derivative can be made to act
equivalently on the $x$-coordinate of $\zeta$. On the other hand,
we can also distort the contour to wrap around $\zeta$, where it
simply shifts the argument of $\Phi$. The result is that we get
exactly the same PDE as in (\ref{pde}), with the identification
$$
g=4/\kappa\,.
$$

Of course this was just one example. Let us see how to proceed
more generally. In radial quantisation, the insertion of the bcc
field $\phi_{2,1}(0)$ gives a state $|\phi_{2,1}\rangle$. Under
the infinitesimal mapping $f_{dt}$ we get the state
$$
\left(1-(2\hat L_{-2}dt+\hat L_{-1}\sqrt\kappa
dB_t)\right)|\phi_{2,1}\rangle\,,
$$
or, over a finite time, a time-ordered exponential
\begin{equation}\label{Texp}
{\bf T}\exp\left(-\int_0^t(\hat L_{-2}dt'+\hat L_{-1}\sqrt\kappa
dB_{t'})\right)|\phi_{2,1}\rangle\,.
\end{equation}
The conformal invariance property of the measure on the curve then
implies that, when averaged over $dB_{t'}$, this is in fact
independent of $t$. Expanding to $O(t)$ we then again find
(\ref{NS}) with $g=4/\kappa$. Since this is a property of the
state, it implies an equivalence between the two approaches for
all correlation functions involving $\phi_{2,1}(0)$, not just the
one considered earlier. Moreover if we replace $\sqrt\kappa dB_t$
by some more general random driving function $dW_t$, and expand
$(\ref{Texp})$ to any finite order in $t$ using the Virasoro
algebra and the null state condition, we can determine all moments
of $W_t$ and conclude that it must indeed be rescaled Brownian
motion.

Of course the steps we used to arrive at this result in CFT are
far less rigorous than the methods of SLE. However, CFT is more
powerful in the sense that many other similar result can be
conjectured which, at present, seem to be beyond the techniques of
SLE. This is part of an ongoing symbiosis between the disciplines
of theoretical physics and mathematics which, one hopes, will
continue.

\section{Further Reading}

The basic reference for CFT is the `big yellow book': {\sl
Conformal Field Theory} by P.~di Francesco, P.~Mathieu and
D.~Senechal (Springer-Verlag, 1996.) See also volume 2 of {\sl
Statistical Field Theory} by C.~Itzykson and J.-M.~Drouffe
(Cambridge University Press, 1989.) A gentler introduction is
provided in the 1988 les Houches lectures by P.~Ginsparg and
J.~Cardy in {\sl Fields, Strings and Critical Phenomena},
E.~Br\'ezin and J.~Zinn-Justin, eds. (North-Holland, 1990.) Other
specific pedagogical references are given below.
\begin{enumerate}
\item[2.1] J.~Cardy, {\sl Scaling and Renormalization in
Statistical Physics}, (Cambridge University Press, 1996.)
\item[3.2] P.~Calabrese and J.~Cardy, {\sl Entanglement entropy
and quantum field theory -- a non-technical introduction},
Int.J.Quant.Inf. {\bf 4}, 429 (2006); arXiv:quant-ph/0505193.
\item[6.1] V.~Pasquier, Nucl. Phys. B {\bf 285}, 162, 1986, and J.
Phys. A {\bf 20}, L1229, 1987; I.~Kostov, Nucl. Phys. B {\bf 376},
539, 1992; arXiv:hep-th/9112059. The version in these lectures is
discussed in J.~Cardy, J.Phys. A {\bf 40},  1427 (2007);
arXiv:math-ph/0610030.
\item[6.2] The basic reference is B.~Nienhuis, in {\sl Phase
transitions and critical phenomena}, vol.~11, p.~1, C.~Domb and
J.L.~Lebowitz, eds. (Academic, 1987.) A slightly different
approach is explained in J.~Kondev, Phys. Rev. Lett. {\bf 78},
4320, 1997.
\item[7.1,7.2] For reviews see V.B.~Petkova and J.-B.~Zuber, {\sl
Conformal boundary conditions and what they teach us}, lectures
given at the Summer School and Conference on Nonperturbative
Quantum Field Theoretic Methods and their Applications, August
2000, Budapest, Hungary, arXiv:hep-th/0103007, and J.~Cardy, {\sl
Boundary Conformal Field Theory} in Encyclopedia of Mathematical
Physics, (Elsevier, 2006); arXiv:hep-th/0411189.
\item[7.3] For reviews on the connection between SLE and CFT, see
J.~Cardy, {\sl SLE for Theoretical Physicists}, Ann. Phys. {\bf
318}, 81-118 (2005); arXiv:cond-mat/0503313, and M.~Bauer and
D.~Bernard, {\sl 2D growth processes: SLE and Loewner chains},
Phys.Rept. {\bf 432}, 115-221 (2006); arXiv:math-ph/0602049.
\end{enumerate}
\end{document}